\documentstyle[amssymb,preprint,aps,psfig]{revtex}
\tighten

\begin{document}
\title{Statics and Dynamics of the 10-state mean-field Potts glass model: A Monte Carlo study}
\author{Claudio Brangian, Walter Kob\footnote{Author to whom correspondence should
be addressed. Present and permanent address: Laboratoire des Verres, Universit\'{e}
Montpellier II, 34095 Montpellier, France} and Kurt Binder}
\address{Institut f\"{u}r Physik, Johannes Gutenberg Universit\"{a}t Mainz, D-55099\\
Mainz, Staudinger Weg 7, Germany}
\maketitle
\begin{center} 16. June, 2001 \end{center}

\begin{abstract}
We investigate by means of Monte Carlo simulations the fully connected
$p$-state Potts model for different system sizes in order to see how the
static and dynamic properties of a finite model compare with the,
exactly known, behavior of the system in the thermodynamic limit. Using
$p=10$ we are able to study the equilibrium dynamics for system sizes as
large as $N=2560$. We find that the static quantities, such as the energy,
the entropy, the spin glass susceptibility as well as the distribution
of the order parameter $P(q)$ show very strong finite size effects. From
$P(q)$ we calculate the forth order cumulant $g_4(N,T)$ and the Guerra
parameter $G(N,T)$ and show that these quantities cannot be used to locate
the static transition temperature for the system sizes investigated. Also
the spin-autocorrelation function $C(t)$ shows strong finite size effects
in that it does not show a plateau even for temperatures around
the dynamical critical temperature $T_D$.  We show that the $N$-and
$T-$dependence of the $\alpha$-relaxation time can be understood by means
of a dynamical finite size scaling Ansatz.  
$C(t)$ does not obey the time-temperature superposition principle for
temperatures around $T_D$, but does so for significantly lower $T$.
Finally we study the relaxation dynamics of the individual spins and
show that their dependence on time depends strongly on the chosen spin,
i.e. that the system is dynamically very heterogeneous, which explains
the non-exponentiality of $C(t)$.

\end{abstract}

\section{Introduction}
\label{sec1}

\noindent
In recent years it has been recognized that the relaxation dynamics of
supercooled liquids and the one of spin glass systems have many properties
in common~\cite{Monasson:1995,Bouchaud:1998,Franz:1998,Mezard:1999a}.
Of particular significance was the observation by
Kirkpatrick, Thirumalai, and Wolynes \cite{Kirkpatrick},
that for certain mean
field spin glasses the equations of motion for the spin-autocorrelation
function $C(t)$ are formally the same as the ones that have been derived
previously for the relaxation of particle density correlation functions
of structural glasses~\cite{Gotze}. That result thus opened the possibility
to find one common description for these two classes of glassy systems:
The spin glasses with quenched disorder and the structural glasses in
which the disorder is not included explicitly in the Hamiltonian.

One of the important classes of spin glasses models is the Potts glass
\cite{Kirkpatrick,Elderfield,Erzan,Gross,Carmesin,Cwilich,Scheucher,DeSantis,Schreider,Rehul,Dillmann,Lobe,Marinari,Hukushima:2000},
which is a
generalization of the Ising spin glass 
\cite{Edwards,Sherrington,Binder:1986,Mezard:1987,Fischer,Stein,Parisi:1992,Young},
where the spins $\sigma_i$ can assume two values $(\sigma_{i}=\pm 1)$, to
the case where each Potts ``spin'' can be in one of $p$ discrete states,
$\sigma_{i}\in \{1,2,\ldots ,p\}$, $p$ being an integer.  Just as the
Potts ferromagnet \cite{Potts,Wu,Zia} is a ``workhorse'' for the statistical
mechanics of phase transitions, since it has served to test many methods
and to exemplify many concepts about the subject, one can expect that the
Potts glass will play a similar role for the study of the glass transition
in systems with quenched disorder or the properties of liquids close
to their glass transition temperature, since the possibility to change
$p$ allows to describe different glass transition scenarios, e.g. from
a continuous transition to a discontinuous one. 

It is important to note  that in the Potts glass it is well established
that for $p>4$ one has both a {\it dynamical} transition at a temperature
$T_D$, where the relaxation time associated with $C(t)$ diverges,
and a {\it static} transition at a temperature $T_0<T_D$, where a glass
order parameter appears discontinuously, accompanied by a kink in the
entropy (as well as in the internal energy).  In contrast to this knowledge, for
the structural glass transition the existence of an underlying static
transition, although proposed long time ago~\cite{Kauzmann,Gibbs}, is still
an open question
\cite{Jackle,Parisi:1987,Binder:1999,coluzzi99,sciortino99,Crisanti}. Thus a study of Potts
glasses should help us to understand also better structural glasses and
hence is a welcome addition to the large efforts made to identify which
structural features distinguish the solid glass from the liquid from
which it was formed 
\cite
{Gotze,Jackle,Parisi:1987,Binder:1999,coluzzi99,sciortino99,Crisanti,Kob:1999,Mezard:1999b,Parisi:2000}.

Besides this interest in the Potts glass as a possible prototype model
for the structural glass transition, it can also be considered as
a coarse-grained model for orientational glasses \cite{Hochli,Binder:1992a,Binder:1998}.
Experimentally these systems are created by random dilution of
molecular crystals, which has the effect that  at low temperatures
the quadrupole moments of the molecules freeze in random orientations
\cite{Hochli}. If the crystal anisotropy singles out $p$ discrete preferred
orientations (e.g. the 4 diagonal directions in a cubic crystal), a Potts glass
model with $p$ states may give a qualitatively correct description of
the system. And, last but not least, the Potts glass model of course
completes our knowledge about the different types of phase transitions
and ordered phases that spin glasses can have 
\cite{Binder:1986,Mezard:1987,Fischer,Stein,Parisi:1992,Young},
which provides an additional motivation for the large activity in this
field \cite{Kirkpatrick,Elderfield,Erzan,Gross,Carmesin,Cwilich,Scheucher,DeSantis,Schreider,Rehul,Dillmann,Lobe,Marinari,Hukushima:2000},

In this work, we use Monte Carlo simulations to study the Potts glass
model. Our goal is to clarify to what extent this interesting and
nontrivial mean field behavior which is known exactly in the thermodynamic
limit, $ N\rightarrow \infty $, can be seen for {\it finite $N$}. In
addition, we want to elucidate the dynamical behavior of the model in
greater detail than has been done so far and thus help to clarify the
reasons for non-Debye relaxation in glassy systems.

In the present paper, we shall first define the model and introduce the
quantities that will be investigated (Sec.~2). In Sec.~3 we summarize
what is known about the static behavior of the model and describes our
pertinent numerical results. Sec.~4 is then devoted to the dynamical
properties in the high temperature phase, i.e. above $T_D$, and the finite
size behavior at $T_D$, while Sec. 5 is concerned with the dynamical
behavior of small systems in the low temperature phase. By following the
relaxation of individual spins we are able to study also the ``dynamical
heterogeneities'' \cite{Sillescu,ediger00} at low temperatures. Sec. 6 finally
summarizes our conclusions.

\section{Model and simulation methods}
\label{sec2}

\noindent
In this section we define the Hamiltonian and the observables that
we consider in this work. Subsequently we give the details on the
simulations.

The Hamiltonian of the $p$-state mean-field Potts glass of $N$ interacting
Potts spins $\sigma _{i}$ $(i=1,2,\ldots ,N)$ that can take $p$ discrete
values $\sigma _{i} \in \{1,2,\ldots ,p\}$ is defined as

\begin{equation}
{\cal H}=-\sum\limits_{i<j}J_{ij}(p\delta _{\sigma_i\sigma_j}-1).
\label{eq1}
\end{equation}

\noindent
The ``exchange constants'' (bonds) $J_{ij}$ are quenched random variables
which we assume to be distributed according to a Gaussian distribution
$P(J_{ij})$

\begin{equation}
P(J_{ij})=\frac{1}{\sqrt{2\pi }(\Delta J)} \exp \Bigg[- \frac{
(J_{ij}-J_0)^2}{2(\Delta J)^2}\Bigg] .  
\label{eq2}
\end{equation}

\noindent
The first two moments $J_0$ and $\Delta J$ are chosen as follows:

\begin{equation}
J_0\equiv \left[ J_{ij}\right] _{av}=\tilde{J}_0/(N-1),\quad 
\tilde{J}_0=3-p,  
\label{eq3}
\end{equation}

\begin{equation}
(\Delta J)^2\equiv \left[ J_{ij}^2\right] _{av}-\left[ J_{ij}\right]_{av}^2
=1/(N-1),  
\label{eq4}
\end{equation}

\noindent
where $[ \ldots ] _{av}$ denotes an average over all realizations of
disordered bonds (while thermal equilibrium averages will be denoted as
$\langle \ldots \rangle $). The scaling of the parameters $J_0,\Delta
J$ with $N$ was chosen such as to ensure a sensible thermodynamic limit
both for the spin glass transition and for the ferromagnetic transition
(which occurs for a certain range of values of $\tilde{J}_{0}$). Note
that for $p=2$ Eqs.~(\ref{eq1}) - (\ref{eq4}) simply reduce to the
Sherrington-Kirkpatrick (SK) model of a spin glass \cite{Sherrington}. We also
mention that the term $\sum\limits_{i<j}J_{ij}$ in Eq.~(\ref{eq1}) is
only included for convenience, since it makes the mean energy of {\it
each system}, i.e. for each realization of the disorder, go to zero for
$T\rightarrow \infty $. For the present choice of the parameter $\Delta
J$, the spin glass transition in the replica-symmetric approximation,
within which one finds a second-order transition for $ p<6$, occurs
at a temperature $T_{s}=1$~\cite{Elderfield}. (Here and in the following we
set Boltzmann's constant $k_{B}\equiv 1$).  While for some choices
of $\tilde{J}_{0}$ the system exhibits a transition to a standard
ferromagnetic phase at a temperature $T_{F}>T_{s}$, for our choice of
parameters $T_{F}$ falls far below $T_{s}$ and hence ferromagnetic order
is of no of interest here \cite{Elderfield,Cwilich}

There exists in addition a second transition to a
different type of spin glass phase (sometimes called ``randomly canted
ferromagnetic phase'') \cite{Elderfield,Gross}, at a transition temperature $T_{2}$
which is given by 

\begin{equation}
T_{2}=(p/2-1)/(1-\tilde{J}_0).
\label{eq5}
\end{equation}

\noindent
For the choice of $\tilde{J}_0$ given in Eq.~(\ref{eq3}) one thus finds
$T_2=1/2$. Hence this choice ensures that at the temperatures of interest
in the present study, $T\geq 0.7$, any effect of this second transition
on physical observables should be negligible.

For defining observables like the magnetization, the glass order
parameter, time-dependent spin autocorrelation functions, etc., it is
useful to choose a representation for the spins that takes into account
the symmetry between their $p$ possible states. This can be achieved by
the so-called ``simplex representation'' \cite{Wu,Zia} in which the $p$
states correspond to $(p-1)$-dimensional vectors $\vec{S}_{\lambda}$
pointing towards the corners of a $p$-simplex, i.e.

\begin{equation}
\vec{S}_{\lambda }\cdot \vec{S}_{\lambda ^{\prime }}=(p\delta _{\lambda
\lambda ^{\prime }}-1)\quad \mbox{with } \lambda ,\lambda ^{\prime }=1,\ldots ,p.
\label{eq6}
\end{equation}

\noindent
In our study we consider static as well as dynamical observables. Static
quantities include the energy per spin, 

\begin{equation}
e=[\langle {\cal H}\rangle ] _{av}/N  ,
\label{eq7}
\end{equation}

\noindent
the spin glass susceptibility $\chi _{SG}$, and the spin glass order
parameter distribution function $P(q)$. For defining a spin glass order
parameter, we follow the standard method used for Potts 
glasses \cite{Scheucher,Dillmann,Hukushima:2000} to consider
two replicas $\alpha$ and $\beta$ of the system, i.e. two systems that
have identical bond configurations, and to make for each of them an
independent Monte Carlo simulation. The order parameter tensor is then
defined as

\begin{equation}
q^{\mu \nu }=\frac{1}{N}\sum\limits_{i=1}^{N}(\vec{S}_{i,\alpha})^{\mu }
(\vec{S}_{i,\beta})^{\nu }\qquad \mu ,\nu =1,2,\ldots ,p-1  
\label{eq8}
\end{equation}

\noindent
In an equilibrium simulation of a finite system with no external fields
that couple to odd moments of the order parameter the symmetry is not broken.
Hence it is useful to consider the root mean square order parameter $q$ 
defined as \cite{Kirkpatrick,Scheucher,Dillmann,Hukushima:2000} 

\begin{equation}
q=\left[\frac{1}{p-1}\sum\limits_{\mu ,\nu =1}^{p-1}(q^{\mu \nu })^{2} \right] ^{1/2} 
\label{eq9}
\end{equation}

\noindent
and by calculating a histogram of $q$, i.e. by taking first the thermal
average and then the average over the disorder, one can estimate the
above mentioned distribution $P(q)$. The second moment of this distribution 
is related to the {\it reduced} spin glass susceptibility $\tilde{\chi}_{SG}$:

\begin{equation}
\tilde{\chi}_{SG}=\frac{N}{p-1}[ \langle q^2\rangle ]_{av} = \frac{N}{p-1}\int\limits_0^1q^2 P(q)dq.  
\label{eq10}
\end{equation}

\noindent
(Below we will discuss the relation between $\tilde{\chi}_{SG}$
and the {\it standard} spin glass susceptibility $\chi_{SG}$, see
Eqs.~(\ref{eq27}) and (\ref{eq27n}).)  If there is a second order
transition to a spin glass phase, $\tilde{\chi}_{SG}$ should show a
divergence at the critical temperature. Further interesting quantities
related to the distribution $P(q)$ are the reduced fourth-order cumulant
\cite{Scheucher,Dillmann,Hukushima:2000} 

\begin{equation}
g_4(N,T)= \frac{(p-1)^2}{2}\Bigg(1+\frac{2}{(p-1)^2}-
\frac{[\langle q^4 \rangle ]_{av}}{[\langle q^2 \rangle]_{av}^2}\Bigg)
\label{eq11}
\end{equation}

\noindent
and a quantity called the Guerra parameter \cite{Guerra} 

\begin{equation}
G(N,T) = \frac{[\langle q^2\rangle^2]_{av}-[\langle q^2\rangle]_{av}^2}
{[\langle q^4\rangle]_{av}-[\langle q^2\rangle]_{av}^2}.
\label{eq12}
\end{equation}  

The reason for introducing these ratios of moment is that they
are useful in the context of finite size scaling analyses of phase
transitions. They are defined such that for $N=\infty$ they are zero
in the disordered phase and nonzero in the ordered phase. In the
finite size scaling limit the curves $g_4(T)$, or $G(T)$, for different
system sizes $N$ should intersect in a common point at the static
phase transition point. In particular, $G$ is a measure for the lack of
self-averaging. 

To study the dynamical properties of the system, we will mainly 
focus on the autocorrelation function of the Potts spins,

\begin{equation}
C(t)=\frac{1}{N(p-1)}\sum\limits_{i=1}^{N}[\langle \vec{S}_{i}
(t^{\prime })\cdot \vec{S}_{i}(t^{\prime }+t) \rangle ]_{av} \quad .
\label{eq13}
\end{equation}

\noindent
Note that in thermal equilibrium $C(t)$ depends only on the time difference $t$,
i.e. it is independent of the second argument $t^{\prime }$ occurring on the right
hand side of Eq.~(\ref{eq13}). In practice, for the Monte Carlo sampling
using the Metropolis algorithm \cite{Binder:1997}, the thermal averaging 
$\langle \cdots \rangle $ is a time averaging over the initial
times $t^{\prime }$.

We have also considered a rotationally invariant order parameter
time-displaced correlation function $C_{RI}(t)$ which is defined as

\begin{equation}
C_{RI}(t)=\left[ \frac{\langle \tilde{q}(t) \rangle}{\langle \tilde{q}(0)
\rangle} \right]_{av}
\label{eq14}
\end{equation}
\noindent

with

\begin{equation}
\tilde{q}(t)= \left[\frac{1}{p-1}\sum_{\mu , \nu =1}^{p-1}
\left(\tilde{q}^{\mu \nu}(t)\right)^2 \right]^{1/2} \qquad
\mbox{and} \quad
\tilde{q}^{\mu \nu}(t)=\frac{1}{N}\sum\limits_{i=1}^N(\vec{S}_{i})^{\mu }(t)
(\vec{S}_{i})^{\nu }(0) .
\label{eq14n}
\end{equation}

Note that $\tilde{q}^{\mu \nu}$ is not the same quantity as $q^{\mu
\nu}$ defined in Eq.~(\ref{eq8}), since the latter involves two replicas
$\alpha$ and $\beta$.  However, for $t \to \infty$ the thermodynamic
averages of the two quantities are the same. This means that in this limit
also $\tilde{q}$ and $q$, from Eq.~(\ref{eq9}), are the same. Furthermore
we mention that the expectation value $\langle \tilde{q}(0) \rangle$
occurring in Eq.~(\ref{eq14}) is equal to $1$, as long as
there is no ferromagnetic ordering of the system. It is important to
realize that $C_{RI}(t\rightarrow \infty )$ is not zero, if $N$ is finite.
This follows immediately from Eq.~(\ref{eq14n}), since for large $t$  the quantity 
$\tilde{q}^{\mu \nu}(t)$ is of order $1/\sqrt{N}$ and $\tilde{q}(t)$, as finite 
sum over
such quantities, is hence positive and also of order $1/\sqrt{N}$. From
Eq.~(\ref{eq10}) one also concludes that $C_{RI}(t\to \infty)$ is of order
$\sqrt{\tilde{\chi}_{SG}/N}$.

In our simulations, we have investigated 5 different system sizes, $
N=160,320,640,1280$, and $2560$. The number of samples used to approximate
the quenched average $\left[ \ldots \right] _{av}$ over the bond disorder
was 500 for $N=160$, $200$ for $N=320$, $100$ for $N=640$ and $1280$ and
between $20$ and $50$ for $N=2560$ (depending on temperature). At not too
low temperatures, $T\geq 1$, the straightforward Metropolis algorithm was
implemented \cite{Binder:1997}, picking spins at random and choosing randomly an
orientation for them as a trial configuration. Depending on the energy
difference $\Delta E$ between the old configuration and the trial
configuration, the trial configuration was always accepted if $\Delta
E<0$, else it was accepted with probability $P=\exp (-\Delta E/T)$. For
temperatures $T<1$ the equilibrium configurations were generated with
the ``parallel tempering'' technique \cite{Hukushima:1996,Kob:2000,brangian_phd}. 
These equilibrium
configurations can both be used to study the static properties of
the model and as starting configurations to study the usual Metropolis
dynamics in equilibrium and thus to calculate $C(t)$,
although at very low temperatures the relaxation is so slow that one
cannot follow the complete decay of $C(t)$ to zero.  The total computing
time (in units of single Pentium II processor with 400MHz) used for
this study was of the order of $10$ years.

\section{Static properties of the 10-state Potts glass model}
\label{sec3}

\noindent
In this section we first discuss the analytic results for the static
properties of the model in the thermodynamic limit. Subsequently we 
compare them with the results of the simulations for finite $N$.

If one calculates the free energy of the model given by Eqs.~(\ref{eq1})
- (\ref{eq4}) with the replica method
\cite{Edwards,Sherrington,Binder:1986,Mezard:1987,Fischer,Stein,Parisi:1992,Young},
(without allowing for replica symmetry breaking), one obtains, depending
on the value of $\tilde{J}_0$ but independent of $p$, either a transition
to a spin glass phase (where the spin glass order parameter $q_0$
is nonzero) or to a ferromagnet (with a spontaneous magnetization
$m_0$). Within this approach and close to a critical temperature $T_s$
the free energy density $f(q_0,m_0)$ can be written as follows \cite{Elderfield},

\begin{equation}
-f(q_0,m_0)/T = 
\frac{1}{2} r^{\prime} \left(1-\frac{T_s}{T}\right) q_0^2+
\frac{1}{6}uq_0^3+
\frac{1}{2}r_m m_0^2+
\frac{1}{6}u_m m_0^4+
\frac{1}{6}u^{\prime}q_0^2m_0^2+\ldots  
\label{eq15}
\end{equation}

\noindent
where with our choice of units $T_s=1$, and $r^{\prime }$, $u$, $r_m$,
$u_m$, $u^{\prime}$ are constants that are of no interest to us here. If
the parameters are chosen such that the transition that occurs at
$T_s$ is to a spin glass phase, it is found that the order parameter
distribution $P(q)$ is a $\delta$-function whose position depends on $T$ 
(we consider here only the case $u>0$):

\begin{equation}
P(q)=\delta(q),\quad \mbox{for } T > T_s  
\label{eq16}
\end{equation}

\begin{equation}
P(q)=\delta (q-q_0),\quad \mbox{with }
q_0=\frac{2r^{\prime}}{u}\left( \frac{T_s}{T}-1\right) ,\quad 
\mbox{for } T<T_s .  
\label{eq17}
\end{equation}

Note that the first term on the right hand side of Eq.~(\ref{eq15}) can
also be interpreted  in terms of the spin glass susceptibility $\chi_{SG}$
as $ \frac{1}{4}\chi _{SG}^{-1}q_{0}^{2}$ \footnote{For the definition of
a proper conjugate field to define the spin glass susceptibility, see
references \cite{Binder:1986,Wortis}}. Thus close to $T_s$ one finds for
$\chi_{SG}$ a Curie-Weiss law:

\begin{equation}
\chi_{SG} = \left[2 r^{\prime }(1-T_s/T)\right] ^{-1},\quad T>T_s.
\label{eq18}
\end{equation}

\noindent
The coefficient $r^{\prime}$ in Eq.~(\ref{eq15}) is given by \cite{Elderfield}
(a result coming from the expansion in $q$ of the free energy close to $T_s$)
 
\begin{equation}
r^{\prime}=\frac{p-1}{2}\left(\frac{T_s}{T}\right)^2 \left( 1+\frac{T_s}{T} \right).
\label{eq22}
\end{equation}

\noindent
We thus find for the susceptibility $\chi_{SG}$ around $T_s$ (remember $T_s=1$ in our
normalization):
\begin{equation}
\chi _{SG}^{-1}=2 (p-1)\left( 1-\frac{T_s}{T}\right), \quad T \approx
T_s .
\label{eq27}
\end{equation}
\noindent
The difference between the
standard spin glass susceptibility and the reduced one defined in Eq.~(\ref{eq10})
is just the factor $(p-1)$, related to the 
susceptibility of a system of non-interacting spins.

It is well known \cite{Kirkpatrick,Gross,Cwilich} that, if one allows for replica
symmetry breaking, the prediction that there is a second order transition
to a spin glass phase remains valid only if $p\leq 4$. For $p>4$, a new
type of first-order transition to a glass phase is predicted to occur at a
temperature $T_0$ which is higher than $T_s$. Although at $T_0$ the order
parameter jumps discontinuously from zero to a value $q_0>0$ there is no
latent heat involved in this transition. Instead of Eq.~(\ref{eq17})
the order parameter distribution acquires now a double-$\delta$ function
structure \cite{Kirkpatrick,Gross,Cwilich}, 
\begin{equation}
P(q)=[1-w(T)] \delta(q)+ 
w(T) \delta (q-q_0(T)),\quad T<T_0 . 
\label{eq19}
\end{equation}
\noindent
with $w(T)=1-T/T_0$ for $T \rightarrow T_0^{-}$.
While it is possible to calculate $q_0$ and $T_0$ analytically for
$p\rightarrow 4^{+} $, see e.g. \cite{Cwilich},
\begin{equation}
q_0 = \frac{2}{7}(p-4)+o(p-4)^2,\quad T_0-T_s\propto (p-4)^2+o(p-4)^3,
\label{eq20}
\end{equation}
\noindent
for integer $p>4$ the correct values for $q_0$ and $T_0$ can be obtained
only numerically \cite{DeSantis}. E.g., for our case of $p=10$ the predicted
values are 
\begin{equation}
T_0=1.1312 \quad \mbox{and }q_0(T_0)=0.452  \quad .
\label{eq21}
\end{equation}
\noindent
In the disordered phase, the internal energy per spin $e$ and entropy
per spin $s$ are given by \cite{Elderfield,DeSantis}
\begin{equation}
e=-\frac{p-1}{2}\frac{T_s}{T},\quad 
s=\ln p- \frac{p-1}{4} \left(\frac{T_s}{T}\right)^2.  
\label{eq23}
\end{equation}
and a high temperature expansion gives \cite{Fischer}
\begin{equation}
\tilde{\chi}_{SG}^{-1}= \left[ 1-\left(\frac{T_s}{T}\right)^2\right] ,
\quad T>T_0.
\label{eq27n}
\end{equation}

\noindent
Within the replica symmetric Ansatz these expressions are correct
for $T>T_s$. If one allows for replica symmetry breaking they hold
only for $T>T_0$. Although no {\it explicit} analytic expression for
$e(T)$ and $s(T)$ are known for $T<T_0$ their value can be calculated
numerically~\cite{DeSantis}.

Finally we mention that within the replica Ansatz (symmetric or broken) neither
 $T_s$ nor $T_0$ depend on the choice of $\tilde{J}_0$ from Eq.~(\ref{eq3}), 
provided that the transition temperature $T_F$ from the disordered phase to the
collinear ferromagnetic phase, discussed in section~\ref{sec2}, is not
above $T_0$~\cite{Elderfield,Cwilich}. $T_{F}$ is given by the following equation,
for arbitrary $p$ \cite{Elderfield}

\begin{equation}
T_F^{-1}=\frac{\tilde{J}_0}{(p-2)} \left[-1+\sqrt{1+2(p-2)/\tilde{J}_0^2}.
\right]  
\label{eq24}
\end{equation}

\noindent
DeSantis {\it et al.} \cite{DeSantis} used $\tilde{J}_0=\frac{1}{2}(4-p)$
in which case $ T_F=T_s=1$, independent of $p$. However, this case
is rather special since then the transition temperature $T_2$ to the
randomly canted ferromagnetic phase, discussed in the Introduction,
coincides with $T_s=1$, as can be seen from Eq.~(\ref{eq5}). For our
choice of $\tilde{J}_0=3-p$ we have instead $T_{2}=1/2$ for all $p$
and $ T_{F}=8/(7-\sqrt{65})\approx 0.531$ for $p=10$. Thus with this
choice we hence make sure that the ferromagnetic fluctuations are still
very small at $T_s=1$, even if the system size is rather small.

We now present our numerical results and compare them to the analytical
predictions that we just discussed. Fig.~\ref{fig1}a shows a plot of
the internal energy versus inverse temperature, over a wide temperature
range ($0.7\leq T\leq 2)$, but still clearly above the temperatures $T_F$
and $T_2$ (remember that $\beta_F \equiv 1/T_F\approx 1.88$, $\beta_2
\equiv 1/T_2=2.0$). Also included are the theoretical predictions for $N
\rightarrow \infty$ obtained within the replica symmetric and one-step
replica symmetry breaking theory, respectively. This figure reveals
unexpectedly large finite size effects over a broad temperature regime in
that, e.g., the energy for $N=640$ coincides with the asymptotic result
only if $\beta \lesssim 0.6$. For $\beta \gtrsim 0.6$ clear deviations
from the asymptotic solution are visible, which are larger for smaller
$N$. The numerical data for finite $N$, in the range accessible to our
work, reveal only a smooth crossover from the regime of the disordered
high temperature phase to the regime of the low temperature glass-like
phase, and no indication of the kink at $\beta_0$, predicted by the
one-step replica symmetry breaking theory, is yet visible. As expected
even for $N \rightarrow \infty$, there is no effect of the dynamical
transition at $\beta _{D}\equiv 1/T_D$ on static quantities like the
energy.

Of course it is also of interest to study how at fixed temperature
the energy $e_N(T)$ converges to its asymptotic limit $e_{\infty
}(T)$. Fig.~\ref{fig1}b shows that the energy difference
$e_N(T)-e_{\infty }\left( T\right) $ scales like $N^{-1}$ both for
temperatures above the static transition temperature $T_0$ and for
temperatures below $T_0$, while at $T_0$ a different law, $\left( \propto
N^{-2/3}\right)$, seems to hold (see inset). (Note that we plot here data
for $T_D$ instead of $T_0$ since we have simulated more system sizes at
this temperature. However, since the two temperatures are so close to
each other this difference should not matter for the system sizes investigated here.)

We have also calculated the temperature dependence of the entropy $s(T)$.
This was done by a thermodynamic integration \cite{Binder:1981} of the free energy $f$:

\begin{equation}
s(\beta)=\beta e(\beta)-
\bar{\beta}f(\bar{\beta})+\int_{\bar{\beta}}^{\beta}e(\beta)d\beta\quad.
\end{equation}

\noindent
We have used $\bar{\beta}=0.5$, a temperature at which our
data is no longer sensitive to finite size effects and thus the
replica solution is valid, so that we can use for the free energy
$\bar{\beta}f(\bar{\beta})=\beta e(\bar{\beta})-s(\bar{\beta})$ the
mean-field value $-9/16-\log(10)$ (see Eq.~(\ref{eq23})). The integral
over $e$ was done by using a spline interpolation of our data, with $180$
points for $N=$160, 320, 640 and $100$ points for $N=$1280, 2560. We
think that alternative methods to calculate $e(T)$, such as re-weighting
techniques or methods to directly sample the density of states \cite{Binder:1997},
would not bring a significant advantage in our case. The results are
shown in Fig.~\ref{fig1}c from which we recognize that $s(T)$ shows
similar finite size effects as the energy $e(T)$.

\mbox{From} Eq.~(\ref{eq23}) we see that the entropy at the static transition is

\begin{equation}
s(T_0)=\ln 10-\frac{9}{4}(T_s/T_0)^2\approx 0.544,\quad
\mbox{i.e. } s(T_0)/s(T=\infty )\approx 0.236.  
\label{eq25}
\end{equation}

\noindent
Thus we see that, while at the static transition temperature $T_0$ the
entropy has decreased to less than a quarter of its high temperature value,
it is clearly nonzero (and nonnegative, of course). In supercooled liquids
one often extrapolates the temperature dependence of the entropy (minus the
vibrational entropy of the crystal) to zero and uses this to define the
Kauzmann temperature $T_K$~\cite{Kauzmann}. If one proceeds in the same way
with the current model to obtain a ``Kauzmann temperature'' $T_{K}$ where
the entropy of the metastable high temperature phase vanishes, one obtains
from Eq.~(\ref{eq23})

\begin{equation}
T_K/T_s=\frac{3}{2}(\ln 10)^{-1/2}\approx 0.9885  
\label{eq26}
\end{equation}

\noindent
which is even below the ``true'' metastability limit $T_{s}=1$ of the
disordered phase, where the (extrapolated) static glass susceptibility
is divergent. (Note that the proximity of $T_K$ and $T_s$ happens
accidentally for $p=10 $. E.g. for $p=5$ the general result \cite{Elderfield}
$T_K/T_s= \left[ \frac{1}{4}(p-1)/\ln p\right] ^{1/2}$ implies
$T_K/T_s\approx 0.7882$.) Also a strictly linear extrapolation of $s(T)$
from $T_{s}$ or $ T_{0} $ down to a temperature where this extrapolation
would vanish does not give a meaningful result, of course.

These results show that the idea to locate the static glass transition
temperature by an extrapolation of the (configurational) entropy
function $s(T)$ in the disordered high temperature phase to $s(T=T_K)=0$
\cite{Kauzmann,Gibbs} can be completely misleading, even for a mean-field model
that does indeed exhibit both a dynamical transition (at $ T_D)$ and a
static transition (at $T_0$). While $T_K$ is always lower than $T_0$,
it does not coincide with the stability limit of the metastable high
temperature phase, and thus lacks any physical significance.  Since for
the case of polymers theories were formulated that suggest that $T_{K}$
is the static glass transition temperature~\cite{Gibbs}, it is interesting
to note that a simulation study of the glass transition in the framework
of the bond fluctuation model found a decrease of $s$ from its
high temperature value to about 1/4 of this value, when $T$ is lowered,
but that subsequently the curve $s(T)$ vs. $T$ bends over and a well-defined $T_{k}$
does not exist \cite{Wolfgardt}. The ``configurational entropy'' estimated
by Gibbs and DiMarzio \cite{Gibbs} was simply the total entropy of their
lattice model of polymers, just as the total entropy of our model shown in
Fig.~\ref{fig1}b. Thus this demonstrates that the calculation of $T_K$
in this way is most likely wrong.  Of course, this ``naive'' way to define
the Kauzmann temperature via the vanishing of the (configurational)
entropy $s(T)$ should not be confused with the approach of defining
a ``complexity''
\cite{Kirkpatrick,Binder:1986,Mezard:1987,Fischer,Stein,Parisi:1992,Young,biroli99}.
In that
approach one determines the number of basins in the {\it free} energy
and defines $T_K$ as that temperature at which this number starts to
become exponentially large.

Fig.~\ref{fig2} shows the reduced spin glass susceptibility as a
function of the squared inverse temperature. This representation is
adapted to the theoretical temperature dependence of this quantity,
see Eq.~(\ref{eq27n}), which predicts at $T^{-2}$-law at high
temperatures.
As expected already from the behavior of the
internal energy, even far above $T_s$ and $T_0$ the convergence to the
thermodynamic limit is rather slow. Unfortunately the analysis of the
finite size behavior of $\tilde{\chi}_{SG}$ is not straightforward, as
we will show in the following. For $T<T_0$ and $N\rightarrow \infty $
Eqs.~(\ref{eq10}) and (\ref{eq19}) imply

\begin{equation}
\tilde{\chi}_{SG}^{-1}=\frac{p-1}{Nq_0^2}\frac{1}{(1-T/T_0)},  
\label{eq28}
\end{equation}

\noindent
since for $T<T_{0}$ our definition for $\tilde{\chi}_{SG}$,
Eq.~(\ref{eq10}), simply picks up a contribution due to the nonzero spin
glass order parameter $q_{0}$. As a result, $\tilde{\chi}_{SG}^{-1}$
for $N\rightarrow \infty $ should follow the straight dashed line
in Fig.~\ref{fig2} that represents the replica-symmetric solution
for all $T^{-2} < T_0^{-2}$, while for $ T^{-2} > T_0^{-2}$,
$\tilde{\chi}_{SG}^{-1}$ simply converges towards the abscissa.
This singular behavior of $\tilde{\chi}_{SG}^{-1}$ is explained further
in the inset, where we have added to $\tilde{\chi}_{SG}^{-1}$ the
term $(T_{s}/T)^{2}$. This sum gives unity for $T>T_0$ and $(T_s/T)^2$
for $T<T_0$, as can be seen from Eq.~(\ref{eq27n}). For very large but
finite $N$, $\tilde{\chi}_{SG}^{-1}$ for $T<T_{0}$ exhibits a Curie-Weiss
type divergence at $T_0$, but the amplitude of this effect is only of
order $1/N$, as can be seen from Eq.~(\ref{eq28}). In order to analyze
the finite size rounding of this singularity, one must consider that
for $N$ finite the $\delta$-functions in Eq.~(\ref{eq19}) are broadened
into peaks of finite height and nonzero width
\footnote{A phenomenological attempt to describe the finite size behavior
for the glass transition of Potts models has been made in \cite{Dillmann}, but
this approach is not followed up here, since it is not consistent with
Eq.~(\ref{eq19}) in the limit to $N\rightarrow \infty$.}. Our simulation
results for $P(q)$, see Fig.~\ref{fig3}, do indeed give evidence that a
second peak at $q_{0}\neq 0$ develops, distinct from the peak at small $q$
that exists also in the high temperature phase. However, the statistical
accuracy of $P(q)$ is not very high due to the well known fact that in the
ordered phase this quantity is not self-averaging
\cite{Binder:1986,Mezard:1987,Fischer,Stein,Parisi:1992,Young},
and the number of realizations of the random couplings that we were able
to study is insufficient to overcome this problem. Hence we are currently
not able to do a proper analysis of the finite size effects of $P(q)$,
see Fig.~\ref{fig3}b, and thus cannot make a finite size analysis of
$\tilde{\chi}_{SG}$.

\mbox{From} Fig~\ref{fig3}a we see that even in the high temperature phase,
i.e.  $T>T_0$, we see a peak in $P(q)$ at a {\it finite} value of
$q$. That this is, however, a finite size effect is demonstrated in
Figs.~\ref{fig3}b and \ref{fig3}c where we plot $P(q)$ for different system sizes
and show the first moment of $P(q)$ as a function
of $N$, respectively. We see that for temperatures well above $T_0$ the first
moment vanishes like $N^{-1/2}$. However, close to $T_0$ this type of
extrapolation would give a {\it finite} value of the moment. If instead
an extrapolation with $N^{-1/3}$ is done, see inset, one finds again as
expected that the moment vanishes.  Note that for the second moment of
$P(q)$ we would have at $T=T_0$ again a scaling of the type $N^{-2/3}$,
as we found for the case of the energy. It is interesting to note that
Parisi \textit{et al.} \cite{Parisi:1993}, making use of replica
symmetry breaking scheme, were able to calculate the finite size scaling
exponents for the Ising spin glass, and obtained that 
$\left[\left\langle q^{k} \right\rangle \right] $ scales like $N^{-k/3}$
at the critical temperature (consequently one has that $e$ scales like
$N^{-2/3}$). It is thus possible that the same kind of scaling holds
also for the Potts glass, although one has to keep in mind that both the
nature of the transition and of the replica symmetry breaking pattern is
different.

While the results shown so far demonstrate a rather encouraging
qualitative consistency between the theoretical predictions and the
numerical data, our results for the fourth order cumulant $g_{4}$,
Eq.~(\ref{eq11}), and the Guerra parameter $G$, Eq.~(\ref{eq12}), are
clearly rather worrisome (Fig.~\ref{fig4}). It is seen that the three
curves for $g_{4}(N,T)$ have a rather well-defined common intersection
point at $T\approx 1.31$, and the three curves for $G(N,T)$ have a rather
well-defined common intersection point at $ T\approx1.24$. Motivated by
the standard knowledge and experience with finite size scaling at second
order \cite{Binder:1997,Binder:1992b} and first order 
\cite{Binder:1997,Vollmayr} phase transitions, such
intersection points are commonly taken as estimates of the transition
temperature 
\cite{Binder:1986,Mezard:1987,Fischer,Stein,Parisi:1992,Young},
However, comparison with the
exact result for $T_0$, Eq.~(\ref{eq21}), shows that these intersection points are
spurious, and hence cannot be taken as accurate estimates of $T_{0}$.
This already is obvious from the data alone, because the two quoted
temperatures are not in mutual agreement. We think that in reality neither
the three curves for $g_{4}(N,T)$ nor those for $G(N,T)$ intersect at a
unique temperature. Given the relatively large error bars of the data,
they only can define temperature intervals $\Delta T_{g4}$, $\Delta
T_{G}$, in which the three intersection points fall.  As $N\rightarrow
\infty $, presumably all temperatures of these intersections converge
(slowly!) towards $T_{0}$. Since $T_{0}$ falls distinctly outside the
above intervals, this method of searching for intersection points,
which is so successful for locating phase transitions in pure systems
\cite{Binder:1997,Binder:1992b,Vollmayr}, is a complete failure here. We emphasize this problem
so strongly, because such techniques are commonly used for studying
phase transitions in systems with quenched disorder \cite{Young}.  Again we
stress that an analytical guidance for the description of the finite
size rounding of first order glass transitions would be very useful.

\section{Dynamical properties in the high temperature phase}
\label{sec4}

In this section we briefly review the theoretical predictions for
the relaxation dynamics of the spins. Subsequently we compare these
predictions with the results from the simulations.

The theoretical results of Kirkpatrick {\it et al.} show that the Potts
glass with $ p>4$ states has a ``dynamical transition'' at a temperature
$T_D>T_0$, where non-ergodicity sets in \cite{Kirkpatrick}. For $T\leq T_D$,
the spin correlation function $C(t)$, defined in Eq.~(\ref{eq13}),
no longer decays to zero but gets stuck for $t\rightarrow \infty $
at a nonzero value $q_{EA}(T)$, with \cite{DeSantis}

\begin{equation}
T_D=1.142,\quad q_{EA}(T=T_D)=0.328 \quad .  
\label{eq29}
\end{equation}

\noindent
The details of this transition from ergodic (for $T>T_D$ where
$C(t\rightarrow \infty )=0$) to non-ergodic behavior (for $T<T_D$), as
well as the time dependence of $C(t)$ for temperatures around $T_D$ are in
fact described by equations \cite{Kirkpatrick} formally analogous to equations
proposed for the structural glass transition by idealized mode-coupling
theory \cite{Gotze}. The qualitative behavior of various quantities
expected for $N\rightarrow \infty $ is sketched in Fig.~\ref{fig5}. Note
that for $T>T_D$ and $T<T_0$ we have $q_0=q_{EA}$. In the temperature
range $T_0<T<T_D$ we have, however $q_0=0$ and $q_{EA}>0$.

In Fig.~\ref{fig6}a we show the time dependence for the spin
auto-correlation function $C(t)$ for $N=1280$ and all temperatures investigated. Here
and in the following we will measure time in units of Monte Carlo Steps
(MCS), i.e. the average number of updates per spin.  Surprisingly we
see that even for this rather large system size there is not
yet any clear evidence for the development of a plateau for temperatures
around $T_D$. Note that in the thermodynamic limit this function should
at $T=T_D$ decay to $q_{EA}$, i.e. the horizontal line. In contrast to
this our system with a finite size is always ergodic, since the free
energy barriers separating the various ``valleys'' in phase space remain
finite at all nonzero temperature. Of course every finite system with
no hard-core interactions is in principle ergodic. However, e.g. in
structural glasses it is found that even a few hundred particles are
sufficient to show a pronounced (effective) ergodic to non-ergodic
transition. Thus it is rather astonishing that for the present model the
finite size effects are so strong that even for $N=1280$ and at $T=T_D$
the existence of a plateau can hardly be seen.

Also the time dependence of $C_{RI}(t)$, Eq.~(\ref{eq14}), shows strong
finite size effects, as can be seen from Fig.~\ref{fig6}b. We see that,
contrary to naive expectation, the long time limit of $C_{RI}(t)$ is not
zero but a finite constant. This constant depends on temperature and
below we will discuss its origin and its dependence on system size in
more detail. At any rate, we see that also this correlation function does
not show a plateau even if $T$ is close to $T_D$ and hence we conclude
that also $C_{RI}(t)$ converges only very slowly to its behavior in the
thermodynamic limit.

In order to discuss the system size dependence of the correlation
functions in more detail we show in Fig.~\ref{fig7} $C(t)$ and $C_{RI}(t)$
for different systems sizes and two temperatures. From Fig.~\ref{fig7}a
we see that, at high temperatures, $C(t)$ shows basically no system size dependence.
For low $T$, however, the relaxation becomes quickly
slower with increasing system size and also the shape of the curve
changes noticeably. But even at the largest system sizes accessible at
this temperature we are not able to see a clear two step relaxation as
one would expect for a sufficiently large {\it but finite} system.

The $N$-dependence is different in the case of $C_{RI}(t)$,
Fig.~\ref{fig7}b.  Here we see that even at high temperatures the
correlation function depends on the system size. This is in agreement
with the arguments given in the context of Eq.~(\ref{eq14}) that
$C_{RI}(t\rightarrow \infty )$ should scale like $1/\sqrt{N}$. That
one actually find this size dependence is shown in the inset of
Fig.~\ref{fig7}b. Instead of studying the function $C_{RI}(t)$ one could
of course try to consider the reduced normalized function $\phi (t)=[
C_{RI}(t)-C_{RI}(t \rightarrow \infty)]/[ C_{RI}(0)-C_{RI}(t \rightarrow
\infty]$. However, also this type of correlation function has its problems
since on one hand the final asymptote $C_{RI}(t\rightarrow \infty )$ is
only known to within a certain statistical error, and on the other hand
it shows finite size effects at high temperatures at {\it short} times,
i.e. where $C_{RI}(t)$ is independent of $N$. In view of these problems
with $C_{RI}(t)$ we will in the following focus on $C(t)$ only. However,
this is not a serious restriction, since in the thermodynamic limit
these two functions should show at low temperatures the same time
dependence anyway.  That this is indeed the case for the simulations
can be inferred from Fig.~\ref{fig7}c where we show a parametric plot of
$C_{RI}(t)$ versus $C(t)$ at $T_D$. We see that with increasing system
size the curves do approach the diagonal, as expected.

We now address the temperature and $N$ dependence of the relaxation
time $\tau$ of the system. One possibility to define $\tau$ is

\begin{equation}
C(t=\tau )=0.2  .
\label{eq30}
\end{equation}

\noindent
Note that although the value 0.2 is somewhat arbitrary, it is a reasonable
choice.  The only important thing is that it is significantly less than
the height of the plateau in the thermodynamic limit, $q_{EA}(T=T_D)$,
cf. Eq.~(\ref{eq29}). (If we would define a time $\tau '$ as $C(t=\tau
')=0.5$, on the other hand, $\tau '$ would be finite also below
$T_D$, and even below $T_0$, until $q_{EA}(T)$ has increased up to
$q_{EA}=0.5$, due to the temperature dependence of the order parameter,
see Fig.~\ref{fig5}a.)

Since  for $N\rightarrow \infty $ the dynamics of the model should be
described by (idealized) mode-coupling theory \cite{Kirkpatrick}, we expect
that $\tau(T)$ shows a power law divergence at $T_D$ \cite{Gotze},

\begin{equation}
\tau \propto (T/T_D-1)^{-\Delta },\quad N\rightarrow \infty  \quad ,
\label{eq31}
\end{equation}

\noindent
where $\Delta$ is an exponent which is non-universal (i.e. model
dependent), but typically not very different from $\Delta \approx
2$. In order to test the validity of Eq.~(\ref{eq31}), one can
plot $\tau^{-1/\Delta}$ for a reasonable trial value of $\Delta $
and look whether the data are compatible with a straight line over a
reasonable range of temperature. If this is the case the extrapolation to
$\tau^{-1/\Delta }=0$ should give an estimate for $T_D$. Fig.~\ref{fig8}
shows that for $\Delta =2.0$ the data is indeed rectified for $1.1\leq
T\leq 1.4$, while outside of this temperature range the curves bend.
Since plots for other reasonable choices of $\Delta$ look quite similar,
the value of $\Delta$ can be estimated only within $\pm 0.5$.  In all
cases it is difficult to use the estimates for $T_D$ for finite $N$
to extrapolate to the value of $T_D$ in the thermodynamic limit since
the $N$-dependence is rather weak and the error bars of $T_D(N)$ are,
due to the mentioned extrapolation, quite substantial. For the case of
$\Delta=2.0$ a dependence of the form $T_D(N)-T_D(\infty) \propto 1/N$
seems, however, to be compatible with the data~\cite{brangian_phd}.

In order to provide a more systematic way of extrapolating the relaxation
times to the thermodynamic limit, we assume that the dynamical finite size scaling
hypothesis \cite{Binder:1992b,Suzuki,Hohenberg} holds and make the Ansatz:

\begin{equation}
\tau =N^{z^\ast }\tilde{\tau}\left\{ N(T/T_D-1)^{\Delta /z^\ast }\right\} 
\text{ for }N\rightarrow \infty \text{ and }(T/T_D-1)\rightarrow 0.
\label{eq32}
\end{equation}

\noindent
The scaling function $\tilde{\tau}(\zeta)$ must obey $\tilde{\tau}(\zeta
\rightarrow \infty )\propto \zeta ^{-z^\ast }$ to recover the proper
thermodynamic limit, i.e. Eq.~(\ref{eq31} ). Using $z^\ast$ as a fit
parameter ($\Delta$ is fixed to 2.0), we thus can try to generate a
master curve from the $\tau(T)$ curves for the different system sizes
$N$. That this is indeed possible if one chooses $z^\ast=1.5$ is shown
in Fig.~\ref{fig9}. From this figure we see that for large arguments the
master curve does indeed show the expected power-law with an exponent
$-z^\ast$ (dashed line). For $T=T_D$ the argument of $\tilde{\tau}$
vanishes and hence we expect a $N$-dependence of $\tau$ of the form
$\tau \propto N^{z^\ast}$ and the inset of Fig.~\ref{fig9} shows that
this is indeed the case.

We mention that Eq.~(\ref{eq32}) has a well-based theoretical foundation
for second order phase transitions 
\cite{Binder:1986,Binder:1992b,Suzuki,Hohenberg}, i.e. the
case in which in Fig.~\ref {fig5} the temperatures $T_0$, $T_D$, and
$T_s$ coincide at a unique critical temperature $T_c$. The diverging
relaxation time is then an immediate consequence of a diverging static
susceptibility, and dynamic finite size scaling is a consequence of
ordinary dynamic scaling \cite{Hohenberg}. E.g., for second order transitions
of mean-field spin glass models one has Eq.~(\ref{eq32}) with $\Delta
/z^{\ast }=\gamma _{MF}+2\beta _{MF}=3$, since the static mean field
exponents of the spin glass order parameter and susceptibility are
$\beta _{MF}=1$ and $\gamma _{MF}=1$, respectively \cite{Binder:1986}. Using
the value of $\Delta =2$ \cite{Binder:1986}, one hence finds for $z^{\ast}$ the
value 2/3. This result could be expected since the standard finite size
scaling result for the critical relaxation in spin glasses {\it with short
range interactions} is $\tau \propto L^{z}$ with $z=4$ in the mean field
approximation \cite{Binder:1986}. (Here $L$ is the linear dimension of the system.)
This result can now be translated into the behavior of infinite range
models at the marginal dimension $d^{\ast}=6$, i.e. at the dimension where
mean field theory becomes valid, via $N=L^{d^\ast }$, which yields $\tau
\propto N^{z/d\ast }=N^{z^{\ast} }$, i.e. $z^{\ast }=z/d^{\ast }=2/3$. This
result is also compatible with numerical simulations \cite{Bhatt}.  However,
the value $z^{\ast }\approx 1.5$ found for the present model is clearly
rather unusual and we are not aware of any analytical estimates for
this exponent.

A further interesting question concerns the asymptotic decay of the
correlation function $C(t)$ towards $q_{EA}$ as $t\rightarrow \infty $
at $ T=T_{D}$. In the context of the structural glass transition the
time regime during which the correlation functions are close to the plateau
is called the ``$\beta$-relaxation'' whereas the decay below the plateau is
called the ``$\alpha$-relaxation''~\cite{Gotze}. Mode-coupling theory predicts
that at $T_D$ the approach to the plateau is given by a a power law, i.e.

\begin{equation}
C(t)-q_{EA}\propto t^{-a},\quad T=T_D  .
\label{eq33}
\end{equation}

A naive way to check for the presence of such a power law is to make a
log-log plot of $C(t)-q_{EA}$ versus time. Fig.~\ref{fig10} shows such
a plot for two relatively large systems (curves with open symbols) and
we see immediately that there is no straight line, i.e. no power law
dependence. However, one must recall that $C(t)$ shows a significant
dependence on $N$ and that the prediction of Eq.~(\ref{eq33}) 
seems to hold only for systems much larger than the ones studied here.
Therefore we have tried to make
an extrapolation, keeping time fixed, of the curves for finite $N$
to determine the curve for $N=\infty$. One possibility for such an
extrapolation is to plot $C(t,N)$ vs. $1/N$. While such a graph gives
a straight line for $N\geq 480$, for smaller $N$ a curvature is clearly
apparent. Therefore we tried whether a plot of $C(t,N)$ vs. $\exp (-const
\cdot N)$, where the constant is a fit parameter, gives a straight
line and found that for a broad range of times (from $t=10$ to $t=300$)
this is indeed possible if the constant is approximately 1/400~\cite{brangian_phd}.

In Fig.~\ref{fig10} we have included the results of these two
extrapolations also and we see that they do not give the same
result. Since {\it a priori} it is not clear which type of extrapolation (if
any!) is the correct one, it is difficult to tell what the shape of $C(t)$
in the thermodynamic limit really is. It is very interesting to note,
however, that the extrapolation with the $1/N$ dependence gives a curve
$C(t,N=\infty)$ which is very well compatible with a power law of the
form given by Eq.~(\ref{eq33}). Thus this gives some evidence that the
extrapolation with $1/N$ is the correct one. The value for the exponent
$a$ we read off is $0.33\pm 0.04$.

It is also interesting to note that the theory predicts a one-to-one
correspondence between the value of $a$ and the exponent $\Delta$ from
Eq.~(\ref{eq31})~\cite{Gotze}. For a given value of $a$ one can use 

\begin{equation}
\frac{\Gamma^2(1+b)}{\Gamma(1+2b)} = \frac{\Gamma^2(1-a)}{\Gamma(1-2a)}
\label{eq33n}
\end{equation}

\noindent
to calculate $b$. (Here $\Gamma(x)$ is the usual $\Gamma$-function.). The
value of $\Delta$ is then given by $1/2a+1/2b$. If one uses the value
$\Delta=2.0$ and the above relations one finds $a=0.36$, in very good
agreement with the value determined from Fig.~\ref{fig10}.

Two other important results from mode-coupling theory concern the 
shape of the correlators close to $T_D$ in the $\alpha$-relaxation regime,
i.e. in the time window where they fall under the plateau. The theory
predicts that in this time regime the correlators can be approximated well
with the Kohlrausch-Williams-Watts function, $\exp(-(t/\tau)^{\beta})$,
a functional form that has been found to work very well in many glassy
systems \cite{Gotze,Jackle,Binder:1999,Kob:1999}. We find, however, that even close to $T_D$
and the largest systems used, this functional form does not give a good fit
to the data.

The second prediction of the theory concerning the $\alpha$-relaxation
is the so-called time-temperature superposition principle. This principle
implies that the correlator $C(t)$ can be written as

\begin{equation}
C(t,T)=\tilde{C}(t/\tau(T)) ,
\label{eq34}
\end{equation}

\noindent
where $\tilde{C}(x)$ is a temperature independent scaling function.
The validity of Eq.~(\ref{eq34}) can be checked if one plots $C(t,T)$
versus $x=t/\tau(T)$. If the superposition principle is valid the
curves for the different temperature should fall on a master curve for
$x\approx 1$ and large $x$. For very small values of $x$, i.e. in the
early $\beta$-regime, no master curve is expected, since Eq.~(\ref{eq34})
is supposed to hold only in the $\alpha$-regime. Fig.~\ref{fig11}
shows this kind of scaling plot and we see that even for a rather large
system, $N=1280$, there is no indication of such a time-temperature
position principle. Of course, it is possible that this failure to verify
Eq.~(\ref{eq34}) is simply due to finite-size effects. Thus, it would be
desirable to check Eq.~(\ref{eq34}) for much larger systems. However,
in view of the strong size effects on the relaxation time $\tau$ near
and below $T_D$, see Fig.~\ref{fig13} below, this is impossible for us
with the present computer resources. 

Although we have just seen that the time-temperature superposition
principle does not hold close to $T_D$ {\it for the accessible system
sizes} it is interesting to see whether this is the case also at lower
temperatures. For $N=160$ we have been able to go to temperatures as low
as $T=0.7$ and in Fig.~\ref{fig12} we show the correlator as a function of
$t/\tau$. If the curves for all temperatures are considered one finds that
the superposition principle does again not hold (main figure). However,
if one uses only the curves for the lowest temperatures, see inset, one
finds that they all collapse onto a nice master curve. Thus we conclude
that at sufficiently low temperatures the time-temperature superposition
principle does indeed hold. We mention also, that the shape of this master
curve is {\it not} an exponential, but that a stretched exponential with
an exponent around 0.43 gives a satisfactory fit.

In order to be able to understand this result a bit better, one needs
to understand in more detail the relaxation of our model for $T<T_D$
and finite $N$, where all free energy barriers in phase space must
obviously be still finite. One could argue that at low temperature the
largest barrier dominates the dynamics and hence the relaxation depends on
temperature only via a temperature dependent prefactor. This temperature
dependence would have to be Arrhenius like and in order to check this we
show in Fig.~\ref{fig13} the $T$-dependence of $\tau$ for the different
system sizes.

\mbox{From} this figure we see that at the lowest temperatures the $T$-dependence
of $\tau$ for the smallest system is indeed Arrhenius like. For
temperatures around $T_D$ and higher, one sees however significant
deviation from this type of temperature dependence. Also for $N=320$ one
can see at the lowest temperature an Arrhenius law, but the activation
energy is significantly larger than the one for $N=160$. Since for
increasing system size the lowest accessible temperature becomes
higher and higher, it is not possible to see anymore the crossover
from the non-Arrhenius $T$-dependence at intermediate temperatures to
the Arrhenius dependence at small $T$. But the plot clearly shows that
at $T_D$ the relaxation times increases quickly with increasing system
size, in agreement with the result from Fig.~\ref{fig9} (inset). Due to
our present inability to equilibrate larger systems also significantly
below $T_D$, we cannot determine reliably the $N$-dependence of the
activation energy of the Arrhenius law found at low temperatures.
We mention that MacKenzie and Young found for the Sherrington-Kirkpatrick
model , i.e. the mean-field Potts glass for $p=2$, that for small
systems ($N\leq 128$) and low temperature ($T=0.6 T_D=0.6T_s$)
the relaxation times increase like $\tau(N) \propto \exp( const\cdot
N^{0.5})$~\cite{mackenzie} whereas in a recent paper Billoire and Marinari
\cite{Billoire} give evidence that the exponent of the power law is 1/3.
If we consider a low temperature, $T=0.7$, this type of $N$-dependence
of $\tau$ is compatible with our data with an exponent 0.5.  However,
if we determine the $N$-dependence of the activation energy in the
temperature regime where $\tau(T,N)$ shows an Arrhenius law, we find
that this energy increases only very slowly, i.e. like $\log(N)$ or a
power of $N$ with a small exponent ($\approx 0.1=1/p$). (Note that the
reason for the two different $N$-dependencies is related to the fact
that the prefactor of the Arrhenius law depends on $N$ also.)

Fig.~\ref{fig13} shows clearly that for small systems, $N=160$, 320,
and 640, it is possible to explore the region of temperatures below both
the dynamical as well as the static transition. Note that in simulations
of models for the structural glass transition \cite{Kob:1999} it has never
been possible to explore such low temperatures for comparable numbers of
particles~\footnote{An exception are simulations of strong glass formers.
E.g. in Ref.~\cite{horbach99} it was shown that the system could be
equilibrated even at temperatures as low as 0.8$T_D$. However, one
was still way above the Kauzmann temperature.}. On the other hand such
models \cite{Kob:1999} do not seem to be much plagued by finite size effects,
although for certain models for structural glasses they have recently
been found~\cite{kim}.

\section{Relaxation of individual spins in the low temperature phase}
\label{sec5}

In the previous section we have investigated the relaxation dynamics of
the {\it whole} system and found that at low temperatures it shows a
non-Debye behavior.  In the present section we focus on the dynamics
of the individual spins in order to obtain a better understanding for
the occurrence of this non-exponentiality.

In recent years it has been recognized that the non-exponential
relaxation in supercooled liquids is often related to the so-called
``dynamical heterogeneities''~\cite{Sillescu,ediger00}. This means that the details of
the relaxation dynamics of the individual particles (relaxation time,
amplitude of the $\alpha$-relaxation, etc.) is different for each
different particle. The reason for this difference is (most likely)
the fact that each particle has a slightly different neighborhood which
thus affects the dynamics of the particle. Note that these differences
are present only on the time scale of the $\alpha$-relaxation $\tau$,
since afterward the particle has changed its neighborhood and hence
its characteristic dynamics. If the dynamics is averaged over a time
much larger than $\tau$, all the particles behave the same.  For spin
glasses this is different, since the disorder is quenched. Hence the
nature of the dynamics of the individual spins is an intrinsic property
of each spin, since each spin is connected to the other ones by a set
of different coupling constants.  For a spin glass with short range
interactions it is therefore not surprising that each individual spin
has a different relaxation dynamics, and this is indeed what has been
found in simulations~\cite{glotzer98}. For spins systems with long range
interactions the presence of such dynamical heterogeneities 
is, however, not that clear, since each spin interacts
with many different ones and hence one might argue that on average the
different spins show the same relaxation dynamics. The goal of this section is to
investigate this point in more detail.

In order to characterize the dynamics of the individual spins we have
calculated the autocorrelation functions $C_{i}(t)$ for spin number $i$:

\begin{equation}
C_{i}(t)=\frac{1}{p-1}\langle \vec{S}_{i}(t')\cdot \vec{S}_{i}(t'+t)\rangle .  
\label{eq35}
\end{equation}

Note that, in contrast to the case of structural glasses, it is here
possible to average the right hand side over different time origins $t'$,
without loosing the information on the identity of the spin. Due to the
single spin nature of the correlation function $C_i(t)$ it is necessary
to make this average over a sufficiently long time in order to obtain
a reasonable statistics. We found that an average over 1000 
 $\alpha$-relaxation times is needed, and therefore
the following results have been obtained only for relatively small
system sizes and $10$ different samples for every temperature
investigated.

In Fig.~\ref{fig14} we show the time dependence of $C_i(t)$ for all
the spins $i=1,\ldots,N$ for three different system sizes $N$. The
temperature is $T_D$, i.e.  the dynamical critical temperature at
which the {\it average} relaxation dynamics, as measured by $C(t)$,
is already strongly non-exponential. From the figure we see that the relaxation
dynamics for the different spins depends strongly on these spins in that,
e.g., they relax to zero on time scales that span more than an order
of magnitude. At a time where the correlation functions have reached
0.5 of their initial value, the width of the range is even higher and
increases rapidly with increasing system size. Furthermore we see from
the figures that the curves for the individual spins seem to occur in
clusters, i.e. that they do not fill the interval between the slowest
and the fastest relaxation in a homogeneous way. Below we will discuss
the reason for this clustering in more detail. 

In Fig.~\ref{fig15} we show the single spin autocorrelation function
at a lower temperature. Comparing these curves with the ones in
Figs.~\ref{fig14}a and~\ref{fig14}b we see that a decrease of $T$ has made the
distribution of the relaxation dynamics even wider. Also the presence
of the clustering of the curves is now much more pronounced. From
Fig.~\ref{fig15} on also recognizes that the shape of the individual
curves is not uniform at all since the ones which decay slowly tend to
be, in the $\alpha$-regime, much steeper than the ones that decay
more rapidly. A more careful analysis shows that these slow spins show
more or less an exponential relaxation whereas, as can already be seen
from the figure, the fast ones show a strong deviation from a Debye law.
Thus we conclude that the non-Debye behavior of $C(t)$ found at low $T$,
see Fig.~\ref{fig12}, is not due to a superposition of Debye laws with
different relaxation times, but the sum of various different processes,
some of which are Debye-like, some of which are not. 

In order to understand the microscopic reason for the presence of these
dynamical heterogeneities a bit better we have investigated to what
extend the relaxation dynamics of an individual spin correlates with other
quantities. For this it is necessary to characterize this dynamics in some
way. As discussed above, the shape of the curves is not at all uniform,
which makes such a characterization rather difficult. Therefore we decided
to neglect all the variations of the shape completely and to characterize
each curve just by the time it takes the spin to decay to a given value.
Therefore we defined two different relaxation times, $\tau_{i}^{(0.4)}$
and $\tau_{i}^{(0.7)}$, via

\begin{equation}
C_{i}(t=\tau_{i}^{(0.4)})=0.4\text{ and }C_{i}(t=\tau_{i}^{(0.7)})=0.7.
\label{eq36}
\end{equation}

In Fig.~\ref{fig16} we show a scatter plot between $\langle e_i
\rangle$, the average energy of spin $i$, and the relaxation time, for
both definitions of $\tau_i$. We see that there is indeed a significant
correlation between the energy and the relaxation time in that spins with
high energy relax faster than the ones with low energy. This result is
very reasonable since a spin that has a low energy will be reluctant
to change its value and therefore to go (with high probability) to a
state with a higher energy. From the figure we also recognize that the
correlation is present for both definitions of $\tau_i$, from which we
conclude that the details of this definition are not crucial.

In order to investigate this point a bit closer we show in
Fig.~\ref{fig17}a a scatter plot of the relaxation time $\tau_{i}^{(0.7)}$
versus $\tau_{i}^{(0.4)}$ for the two temperatures. We see that
although the correlation is not perfect, it is still very significant
and therefore we conclude that the salient features of the correlation
between the relaxation time and the mean energy shown in Fig.~\ref{fig16}
will be observed even if a more careful characterization of the relaxation
dynamics is made.

Of further interest is the question how the relaxation time of a given
spin at a given temperature is related to the relaxation time of the
same spin at a different temperature. This dependence is related to
the question of ``chaos in temperature'', i.e. how the properties of a
system change if temperature is changed. For mean field type system it is
expected that these dependencies are rather weak~\cite{kondor89,ritort94}.
In agreement with this expectation we find that indeed the relaxation
times $\tau_i$ for $T=0.9$ are strongly correlated with those at
$T=1.142$, see Fig.~\ref{fig17}b, irrespective of the definition
of $\tau_i$. Thus we see that this property seems not to be strongly
affected by finite size effects. In passing we also mention that the mean
energies $\langle e_i(T)\rangle$ between the two temperatures are even
much stronger correlated than the relaxation times~\cite{brangian_phd}.

Before we end we come back to the observation discussed above that some
of the single spin autocorrelation functions occur in clusters (see
Fig.~\ref{fig15}). One potential reason that the relaxation dynamics
of two spins is similar is that they are coupled together strongly,
i.e. that their interaction constant $J_{ij}$ is large. In order to
test this idea we identified for each realization of the disorder those
spins that formed at $T=0.9$ the cluster that relaxed slowest. (This
identification was done visually by means of plots like the one shown
in Fig.~\ref{fig15}b). Say that this cluster involved $k$ curves. We
then determined the $k(k-1)/2$ interaction constants between these $k$
spins. The values of these constants are shown in Fig.~\ref{fig18}
for ten different realizations of the disorder (filled circles). Also
included in the figure is the Gaussian distribution of the coupling
constants given by Eq.~(\ref{eq2}). From this figure we see that most of
the points corresponding to the couplings $J_{ij}$ are to the right of
the mean of the distribution (vertical dashed line). Hence we conclude
that the spins that form the slow cluster of relaxation curves are
coupled together stronger than two arbitrary spins and therefore form a
``dynamic entity''. We note, however, that the fact that two spins are
strongly coupled does not necessarily make them slow~\cite{brangian_phd}
which shows that such a strong coupling is only a necessary but not a
sufficient condition for a slow dynamics.

It is clear that the observations presented in this section are only modest
first steps addressing the dynamics of the individual spins in the low
temperature phase. It certainly would be interesting and useful to
understand better how the distribution of the relaxation times of the
spins depends on the system size and on temperature in order to obtain
a better comprehension how the mean relaxation dynamics of the system
is related to the one of the single spins. However, due to the high
computational demand for this kind of investigations such studies have
to be left to the future.

\section{Conclusions}

\begin{figure}[h]
\psfig{figure=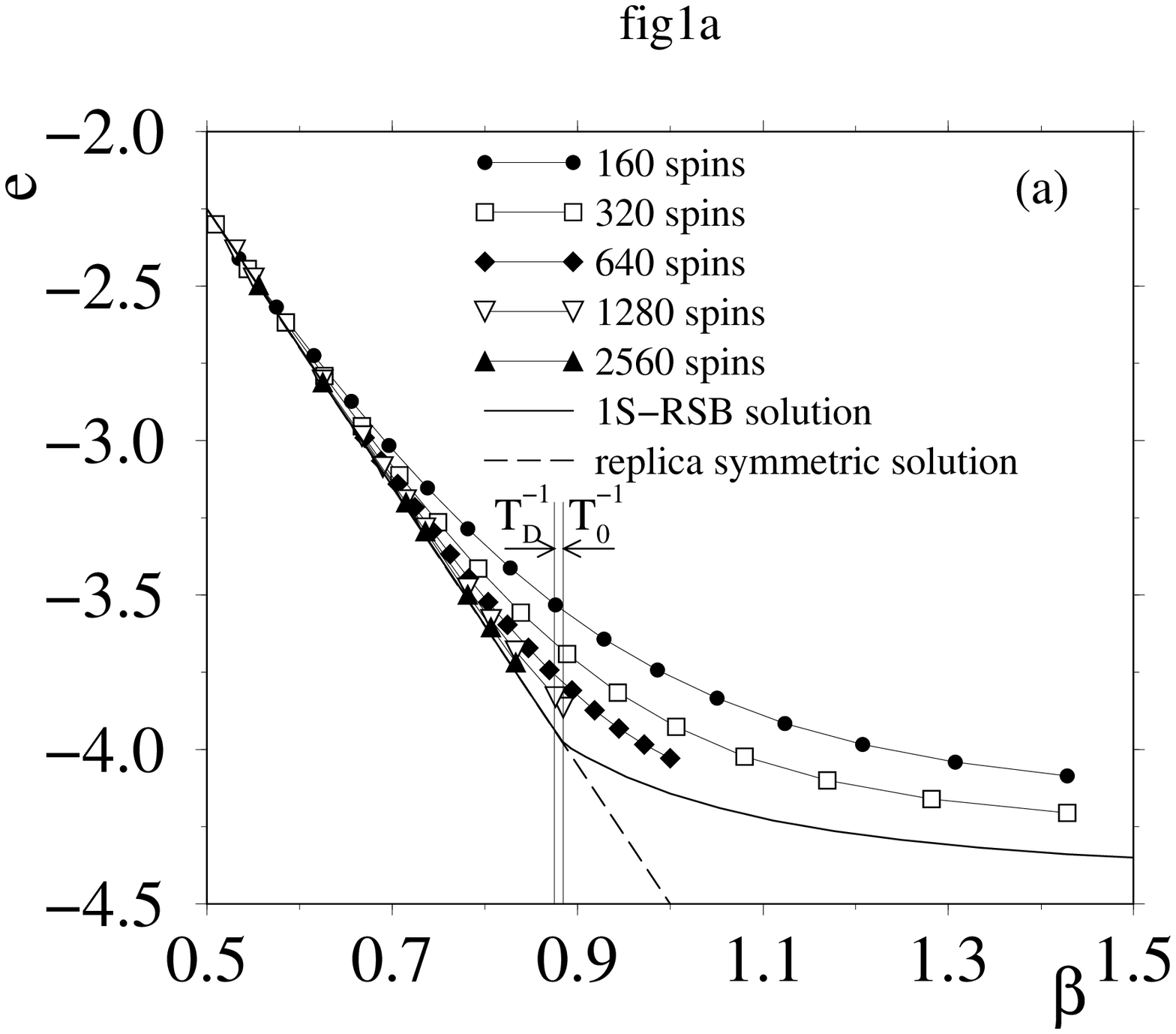,width=10.0cm,height=9.5cm}
\psfig{figure=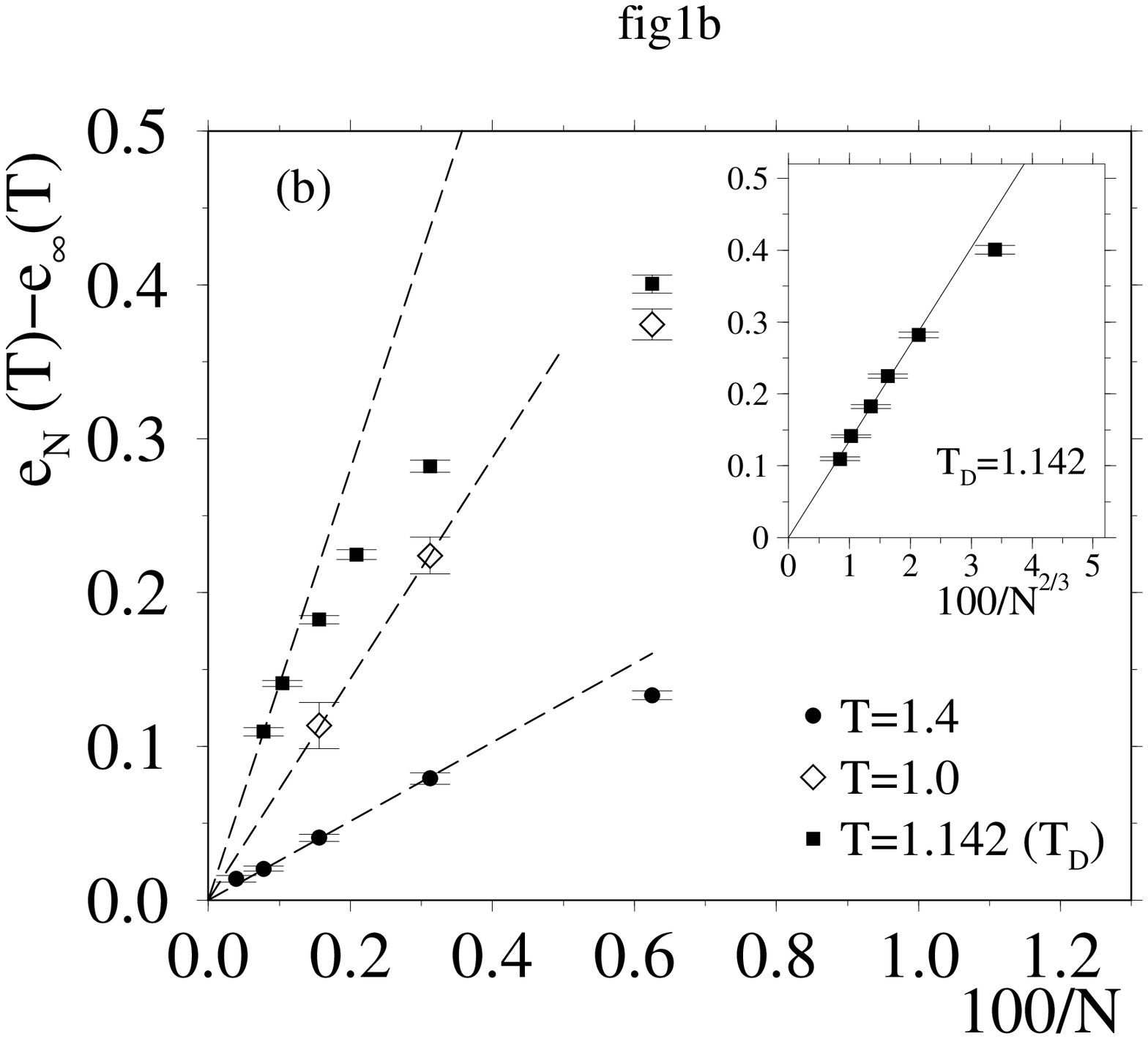,width=10.0cm,height=9.5cm}
\psfig{figure=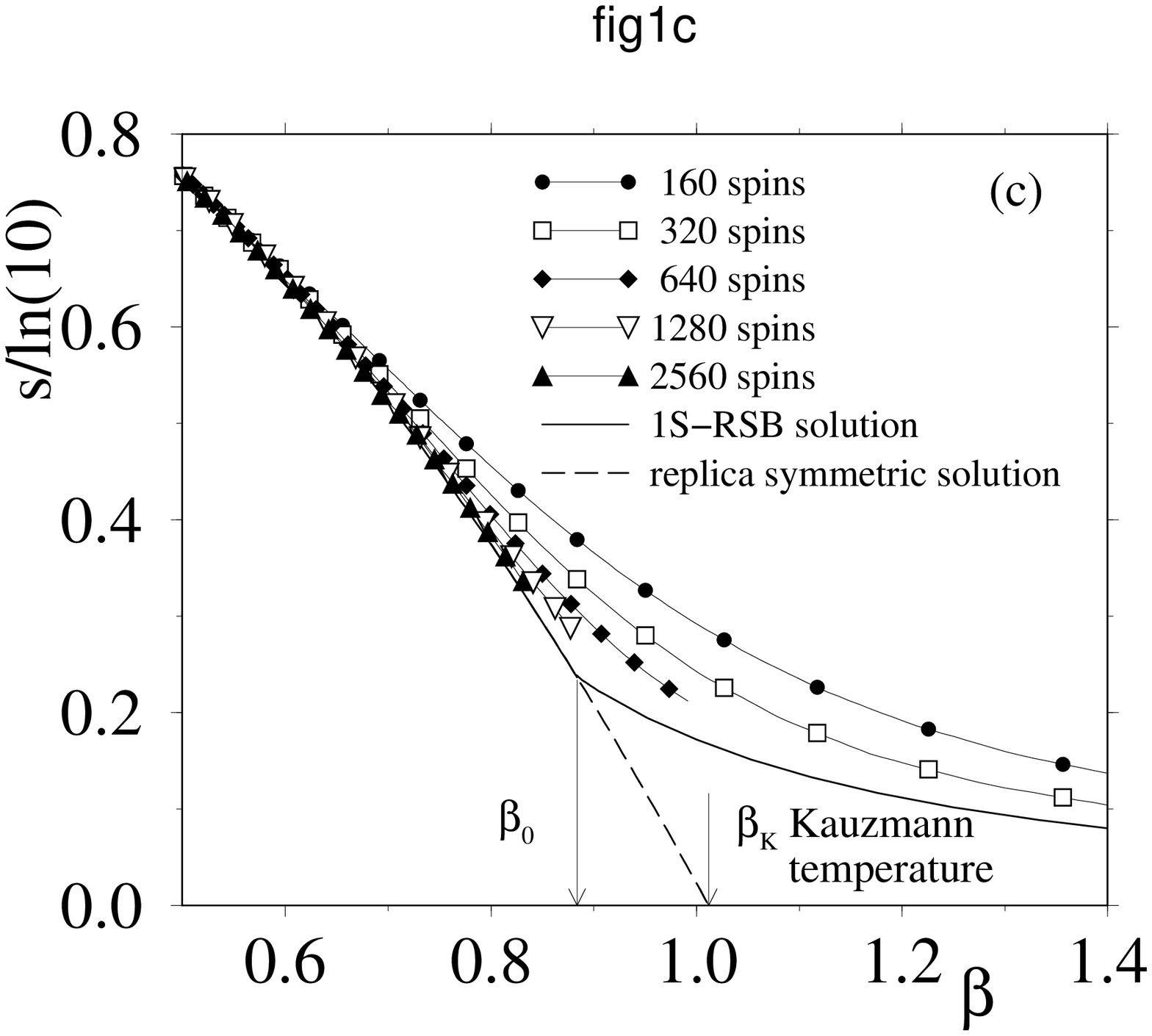,width=8.0cm,height=7.5cm}
\vspace*{-3mm}

\caption{(a) Energy $e$ per spin plotted vs. inverse temperature $\beta
=1/T$ for different system sizes (curves with symbols).  The bold
solid curve shows the one step replica symmetry breaking solution of
DeSantis et al.~ \protect\cite{DeSantis,Ritort_comm}, the broken curve - which
coincides
with the former for $T\geq T_0$ - is the replica symmetric solution,
Eq.~(\ref{eq23}). The thin vertical lines indicated the inverse temperatures
$\beta_D$ (left) and $\beta_0$ (right) of the dynamical transition and
the static transition, respectively.
(b) Analysis of the size dependence of the energy difference
$e_N(T)-e_{\infty }(T)$, using the one-step replica symmetric solution
of DeSantis {\it et al.} \protect\cite{DeSantis} to calculate $e_{\infty
}(T)$. Inset shows the data of $T=T_D=1.142$ plotted vs. $N^{-2/3}$
instead of $N^{-1}$.
(c) Entropy $s$ per spin, normalized by its high temperature value,
plotted vs. inverse temperature for different system sizes (curves
with symbols). The bold dashed and the bold solid curve is the replica
symmetric solution and the one-step replica symmetry breaking solution,
respectivley. Vertical arrows indicate the static inverse transition
temperature $\beta_0$ and the inverse of the ``Kauzmann temperature''
$\beta_K$, where the entropy of the replica symmetric solution vanishes.}
\label{fig1}
\end{figure}

In this paper, we have presented the first detailed Monte Carlo
investigation of the 10 state mean field Potts glass model, for finite
systems with sizes in the range from $N=160$ to $N=2560$ spins. In the
thermodynamic limit, it is known that this model exhibits both a dynamical
transition at $T_D$ where the system stops to be ergodic, and a static
transition at $T_0 < T_D$ where a glass order parameter $q$ appears
discontinuously (Fig.~\ref{fig5}a). The static spin glass susceptibility
$\chi_{SG}$ remains finite both at $T_0$ and $T_D$. It would diverge
only at a still lower ``spinodal'' temperature $T_s<T_0$, if one were
able to follow the disordered branch of the free energy at temperatures
less than $ T_0$. A further relevant temperature is the ``Kauzmann''
temperature $T_K$, Fig.~\ref{fig1}c, defined from the condition
that the entropy of the high temperature phase vanishes. The spin
autocorrelation function $ C\left( t\right) $ decays with time $t$ at
$T\gtrsim T_D$ in a two-step process, and the lifetime $\tau $ of the
``plateau'' diverges as $\tau \propto \left( T-T_D\right)^{-\Delta}$,
in the thermodynamic limit $ N\rightarrow \infty $ (Fig.~\ref{fig5}a). This
behavior entails all qualitative features that one expects to be present
at the structural glass transition (note that the equation of motion
for $C\left( t\right) $ indeed is described by the popular idealized
mode-coupling theory proposed for the structural glass transition!).

The questions asked in the present paper hence are as follows: can we verify
these predictions from Monte Carlo simulations? How are the various
transitions modified (i.e., rounded off) by the finite size of the
considered model systems? Answers to these questions are not only of
interest for a better understanding of the statistical mechanics of the
present model system, they may also be useful to help with the
interpretation of simulations of models for the structural glass transition.
First of all, in the latter case the very existence of the various
temperatures $T_D$, $T_0$, $T_s$, $T_K$ is still open to doubt.
Secondly, even if one believes these temperatures should exist, their
location for a particular model is still uncertain, unlike the present case
where we have so much guidance from the exact solution. Of course, it is
clear that a mean-field model is a rather special limit, and the sharpness of
the dynamical transition at $T_D$ probably is replaced by a smooth
crossover from rather fast relaxation to very slow relaxation, as soon as
one allows the interactions to become short ranged. In this sense, the finite
mean-field Potts glass (where the singularity at $T_D$ is rounded by the
finite size of the system) may be qualitatively similar to the finite range
model (although one should not push this analogy too far).

\begin{figure}[h]
\psfig{figure=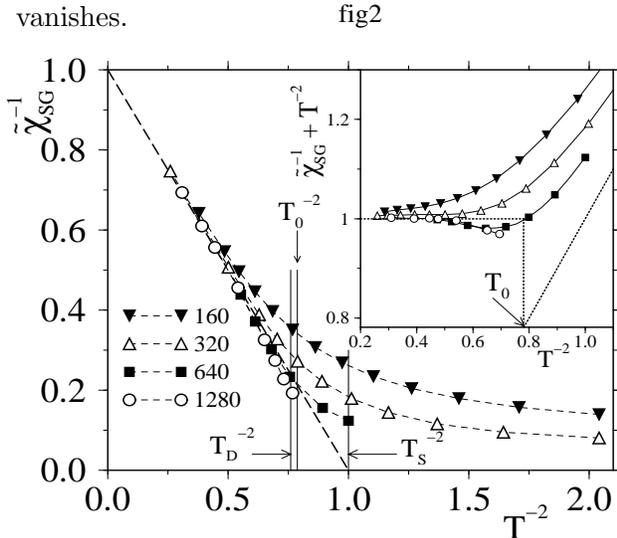,width=8.0cm,height=7.5cm}
\vspace*{-3mm}

\caption{Inverse of the reduced spin glass susceptibility
$\tilde{\chi}_{SG}$ versus the square of inverse temperature for different
system sizes (curves with symbols).  The solid line shows the result of
the replica-symmetric theory, Eq.~(\protect\ref{eq27n}).  Inset: Plot
of $\tilde{\chi}_{SG}^{-1}+(T_s/T)^2$ to illustrate the nonmonotonic
convergence towards Eq.~(\ref{eq27n}). See main text for more details.}
\label{fig2}
\end{figure}

\begin{figure}[h]
\psfig{figure=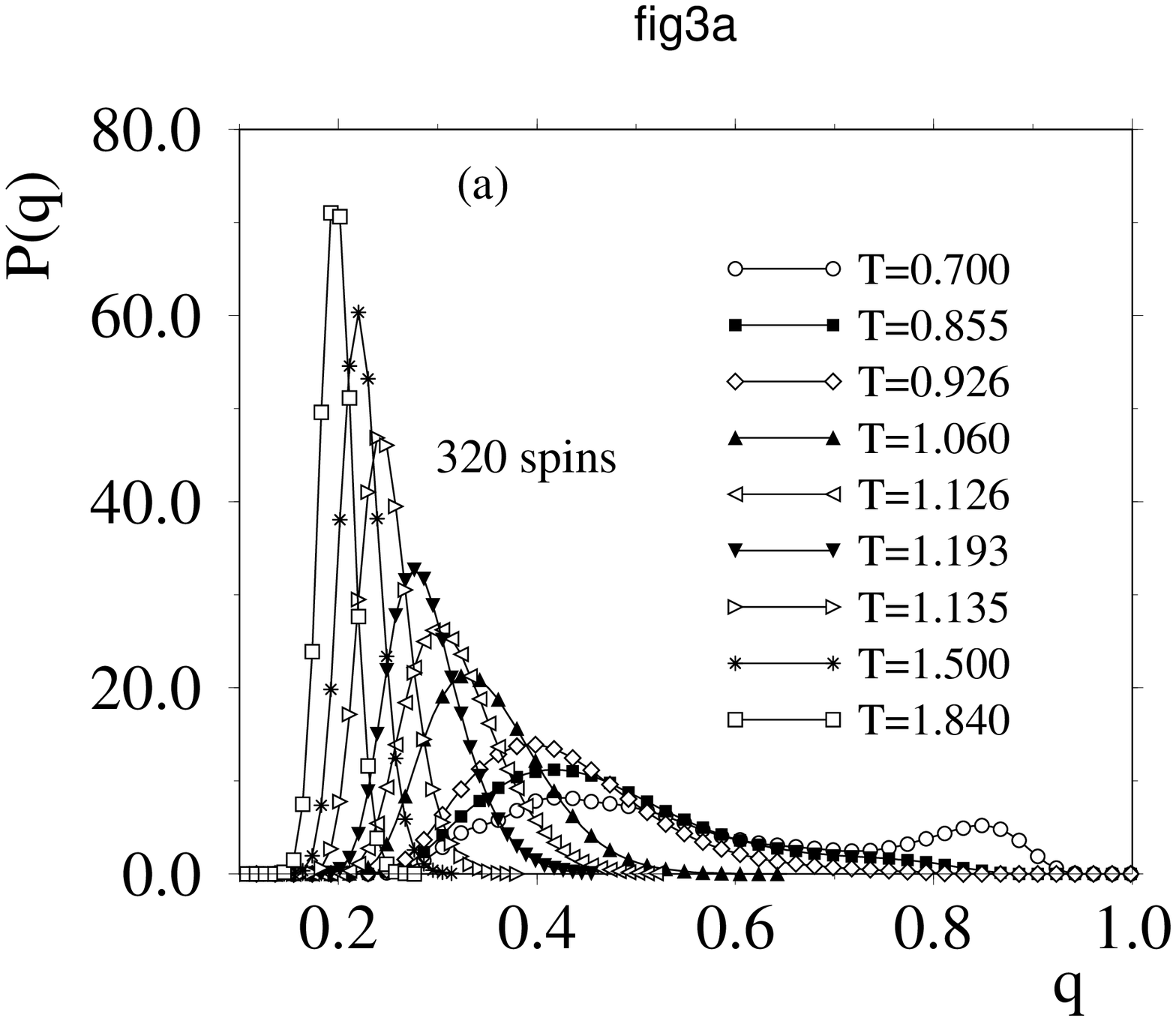,width=10.0cm,height=9.5cm}
\psfig{figure=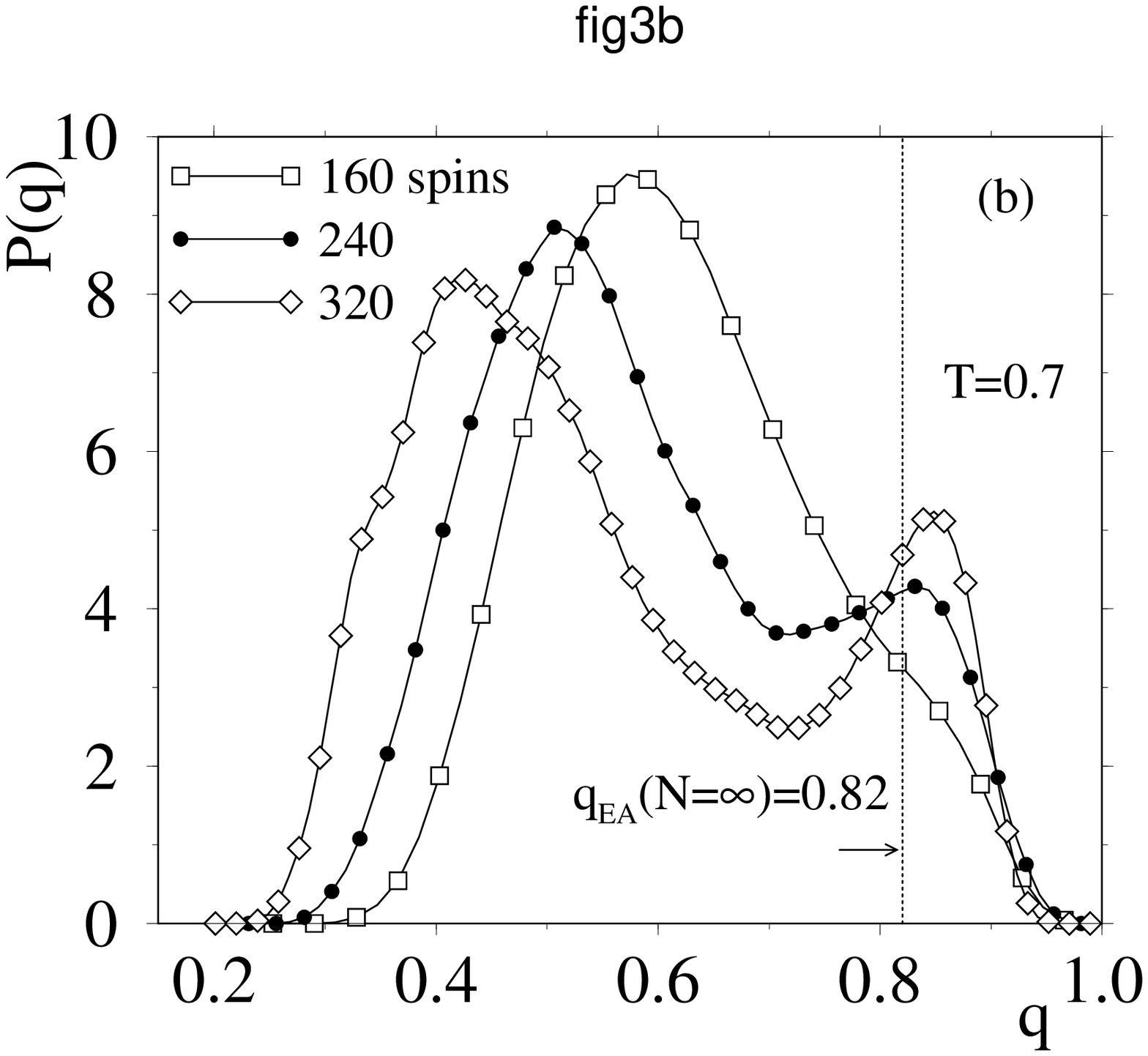,width=10.0cm,height=9.5cm}
\psfig{figure=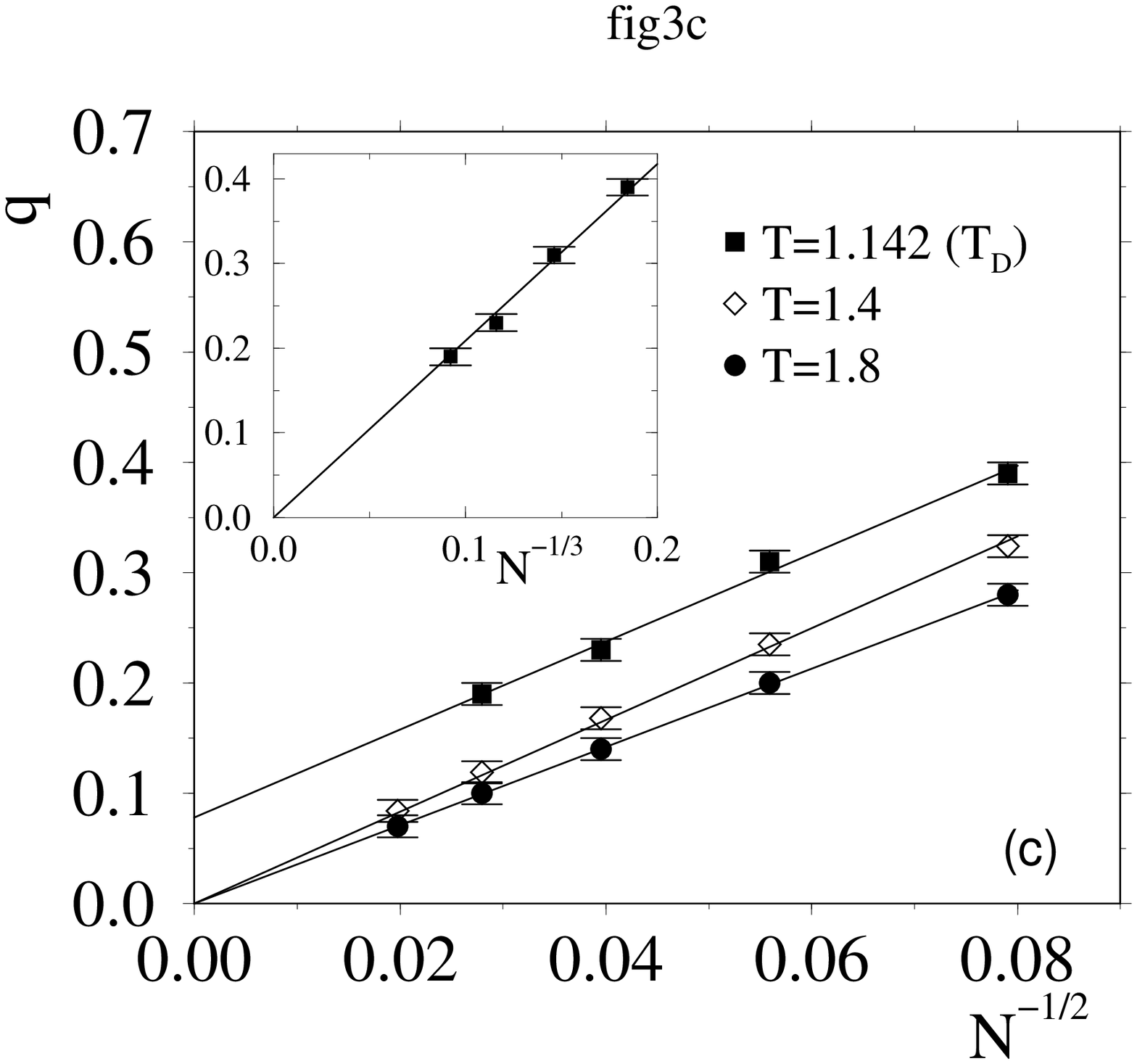,width=10.0cm,height=9.5cm}
\vspace*{-3mm}

\caption{(a) Order parameter distribution $P(q)$ versus $q$ for $
N=320$ and various temperatures. (b) Order parameter
distribution $P(q)$ versus $q$ for $T=0.7$ and the three system sizes
$N=160,240$ and $320$. The asymptotic value of the order parameter (from
Ref.~\protect\cite{Ritort_comm}) is included by a vertical line.
(c) Value of the first moment $\int q P\left( q\right) dq$ of the
order parameter distribution vs. $N^{-1/2}$. The inset shows that close to the
transition temperature $T_0\approx T_D$ this moment scales like $N^{-1/3}$.}
\label{fig3}
\end{figure}

\begin{figure}[h]
\psfig{figure=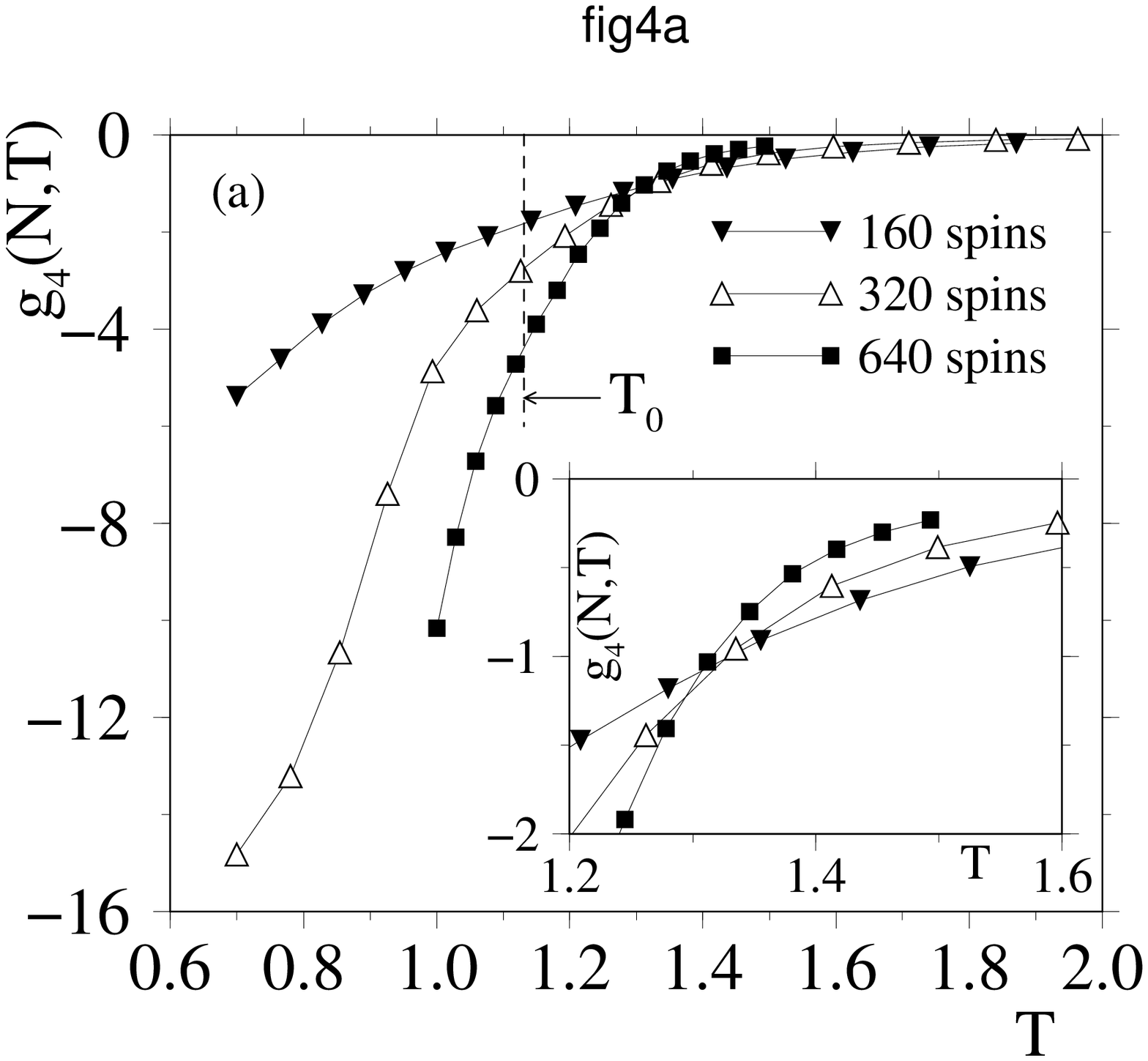,width=10.0cm,height=9.5cm}
\psfig{figure=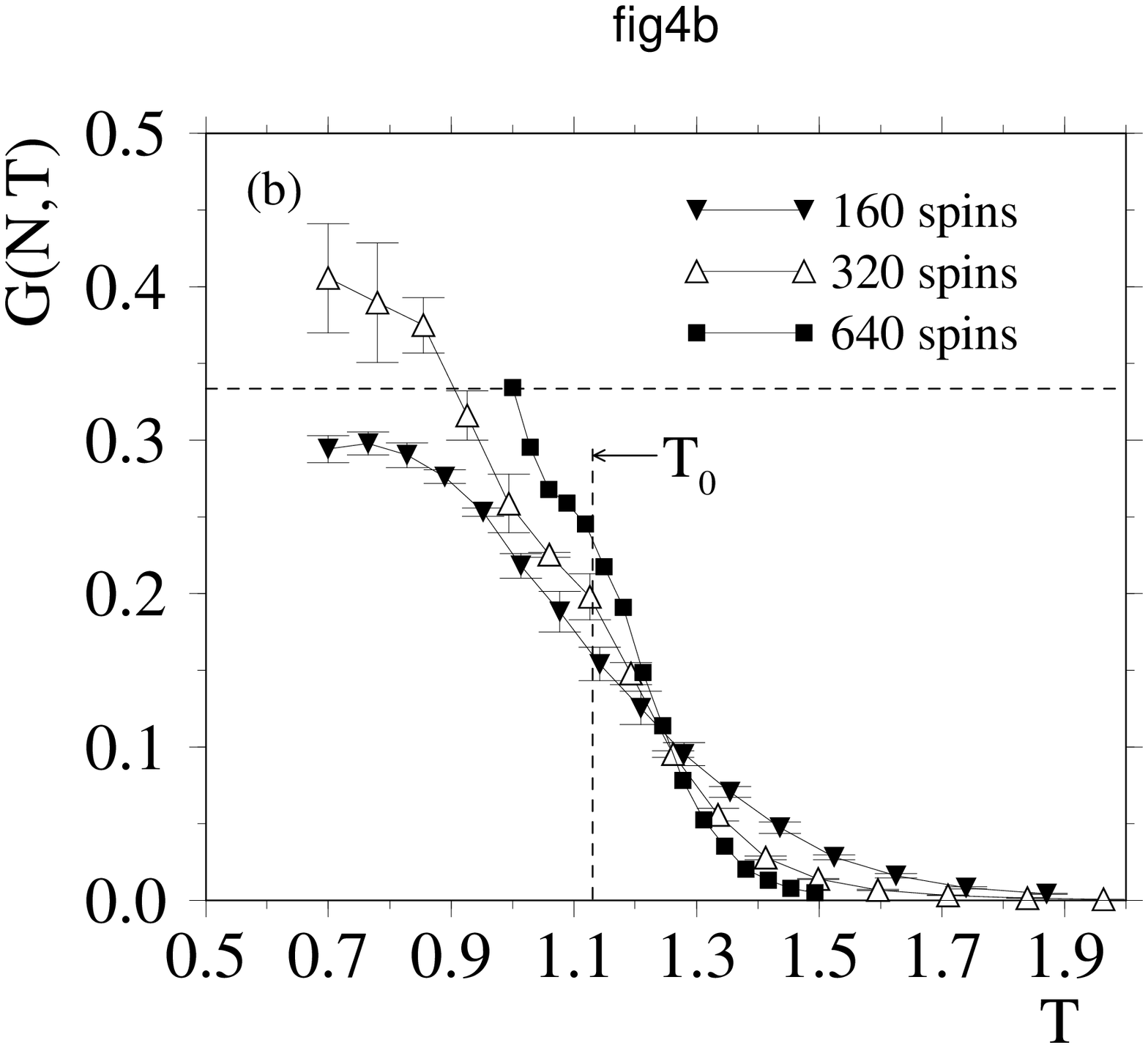,width=9.0cm,height=8.5cm}
\vspace*{-3mm}

\caption{(a) Fourth order cumulant $g_{4}$ plotted versus temperature, for
three values of $N$, $N=160,320$ and $640$. The
vertical straight line highlights the predicted static transition
temperature $T_{0}$. (b) Same as (a) but for the Guerra parameter. The horizontal
dashed line is the theoretical expectation for $T<T_0$.}
\label{fig4}
\end{figure}

\begin{figure}[h]
\psfig{figure=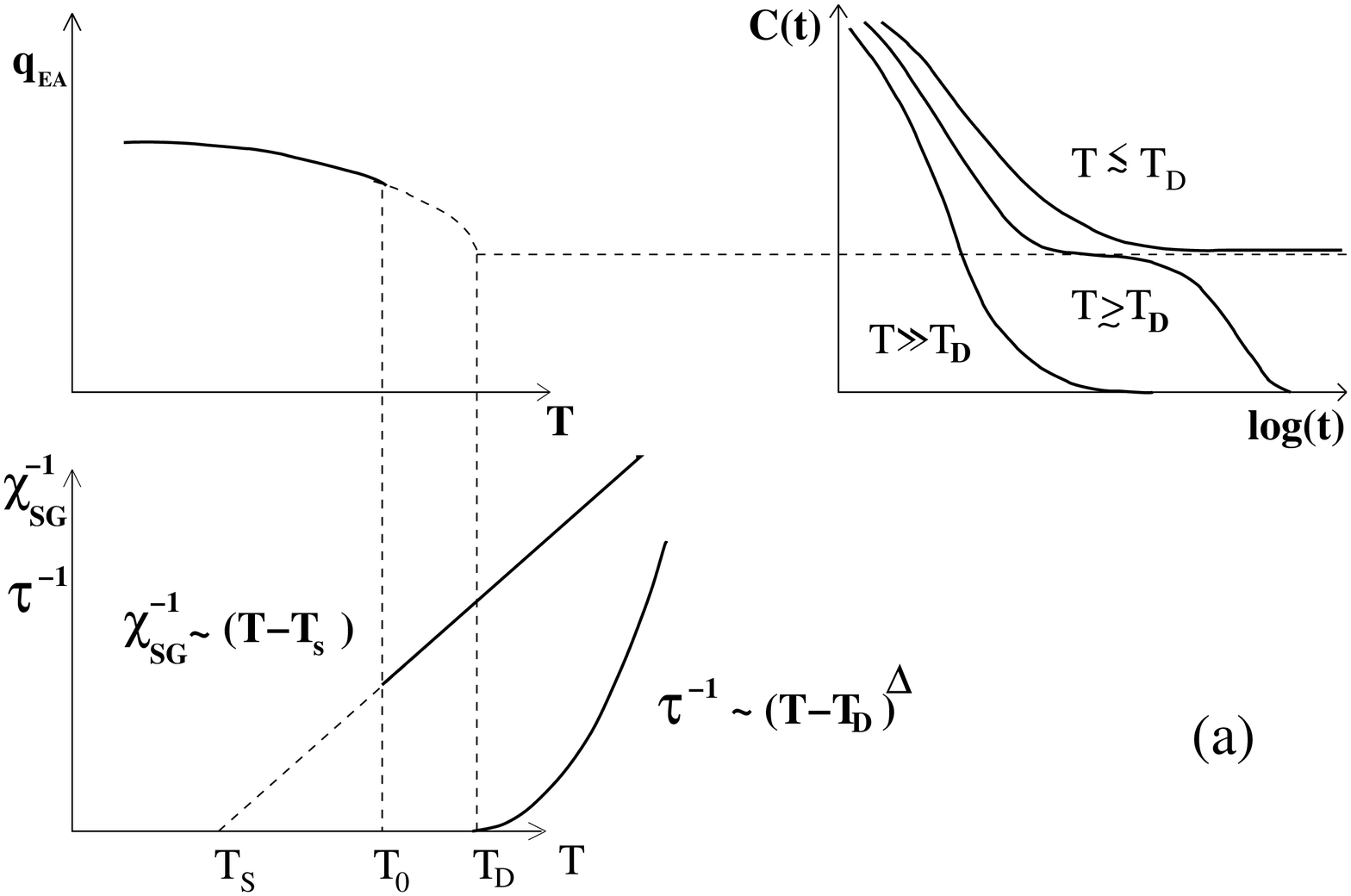,width=14.0cm,height=9.5cm}
\psfig{figure=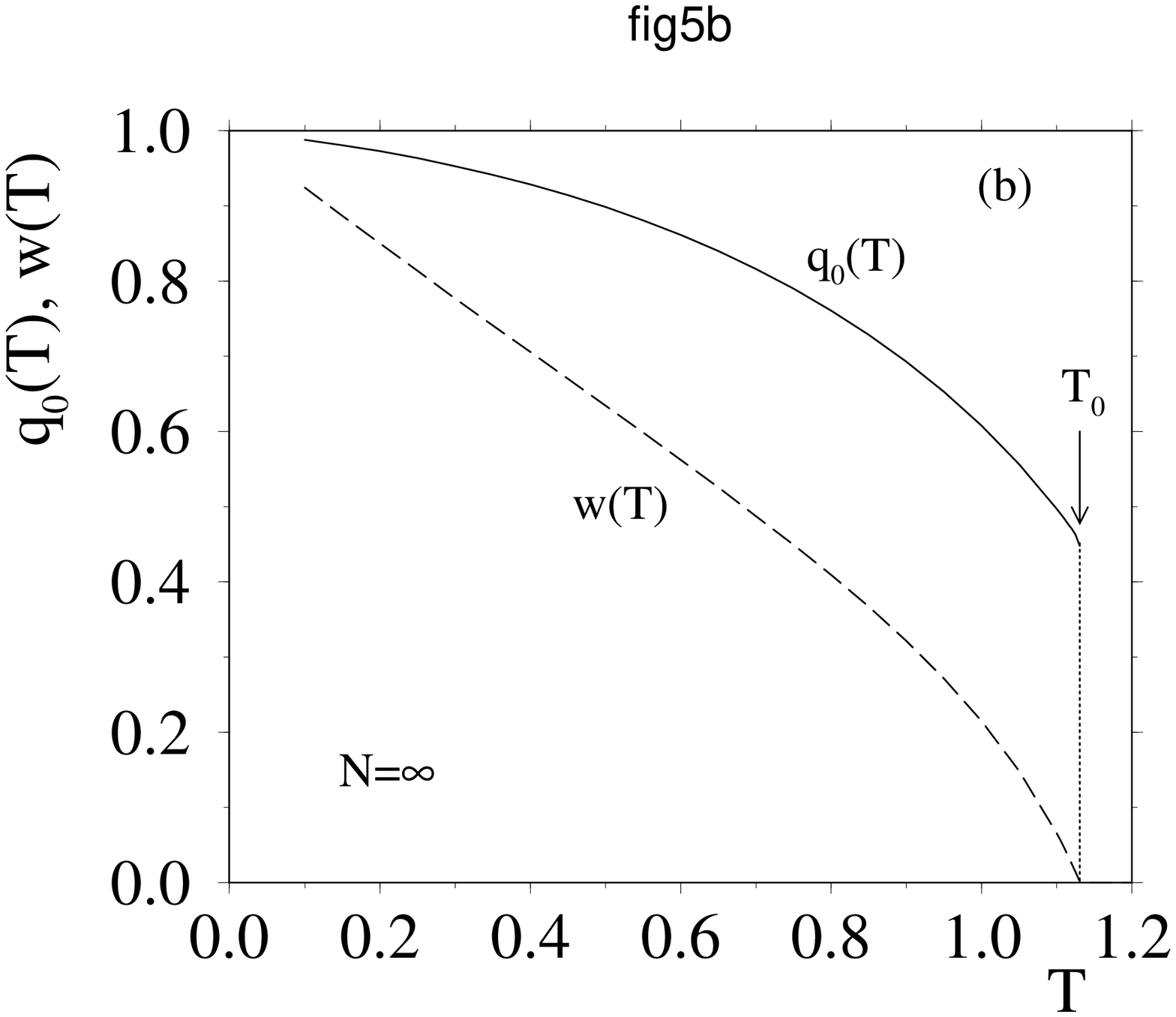,width=8.0cm,height=7.5cm}
\vspace*{-3mm}

\caption{(a) Qualitative sketch of the mean-field predictions for the
$p$-state Potts glass model with $p>4$. The spin glass order parameter,
$q_{EA}$, is nonzero only for $T<T_0$ and jumps to zero discontinously
at $T=T_0$. The spin glass susceptibility $\chi_{SG}$ follows a
Curie-Weiss-type relation with an apparent divergence at $T_s<T_0$, see
Eq.~(\protect\ref{eq27}). The relaxation time $\protect\tau $ diverges
already at the dynamical transition temperature $T_D$. This divergence
is due to the occurence of a long lived plateau of height $q_{EA}$ in
the time-dependent spin autocorrelation function $C(t)$ . From Brangian
{\it et al.} \protect\cite{Brangian}. (b) Temperature dependence of $q_0$ and
$w(T)$  (see Eq.~(\ref{eq19})) for $p=10$, as obtained from the one-step
replica symmetric solution of DeSantis {\it et al.} \protect\cite{DeSantis}.}
\label{fig5}
\end{figure}

\begin{figure}[h]
\psfig{figure=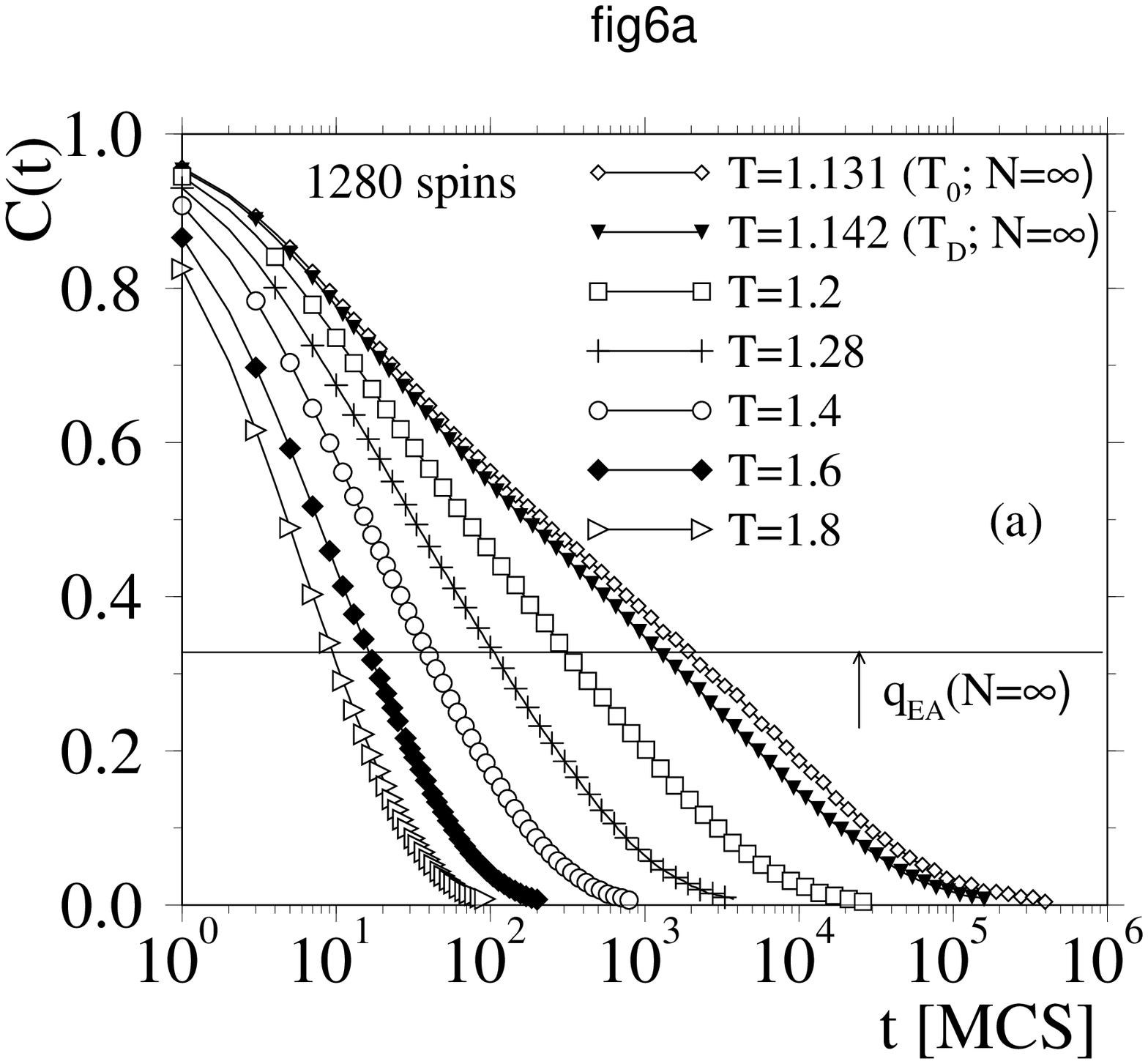,width=10.0cm,height=9.5cm}
\psfig{figure=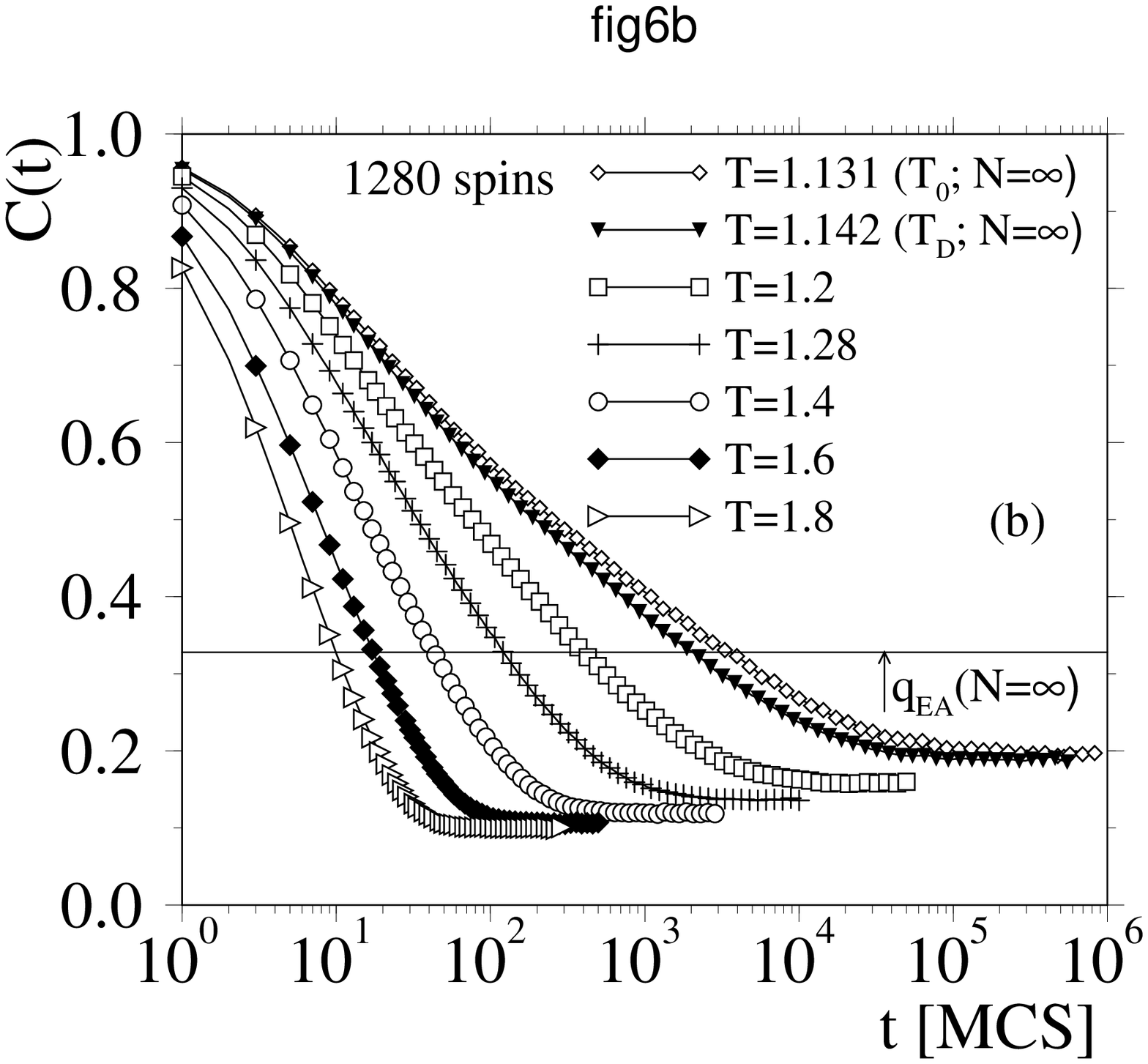,width=10.0cm,height=9.5cm}

\caption{(a) Time dependence of the correlation function $C(t)$,
Eq.~(\ref{eq13}), for $N=1280$ and various temperatures.  Also included
is data for the predicted values of the static, $T_0$, and dynamic, $T_D$,
transition temperature. The horizontal straight line shows the theoretical
prediction from Ref.~\protect\cite{DeSantis} for the Edwards-Anderson order
parameter at $T_D$, $q_{EA}= \mathrel{\mathop{\lim }\limits_{t\rightarrow
\infty }} C(t)$, cf.~Eq.~(\ref{eq29}). (b) Same as (a) but for the
rotationally invariant correlation function $C_{RI}(t)$ defined in
Eq.~(\ref{eq14}).}
\label{fig6}
\end{figure}

\begin{figure}[h]
\psfig{figure=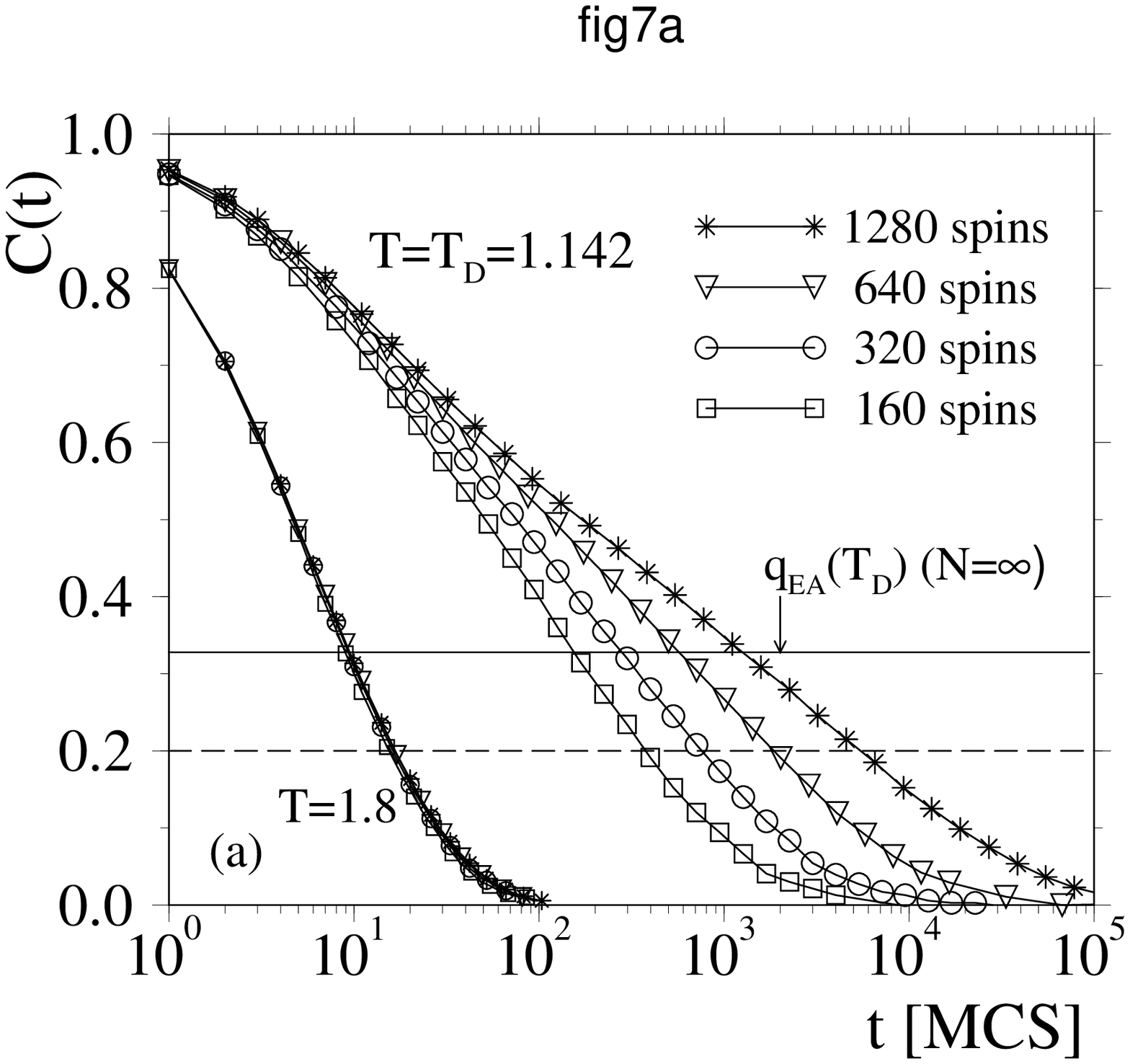,width=10.0cm,height=9.5cm}
\psfig{figure=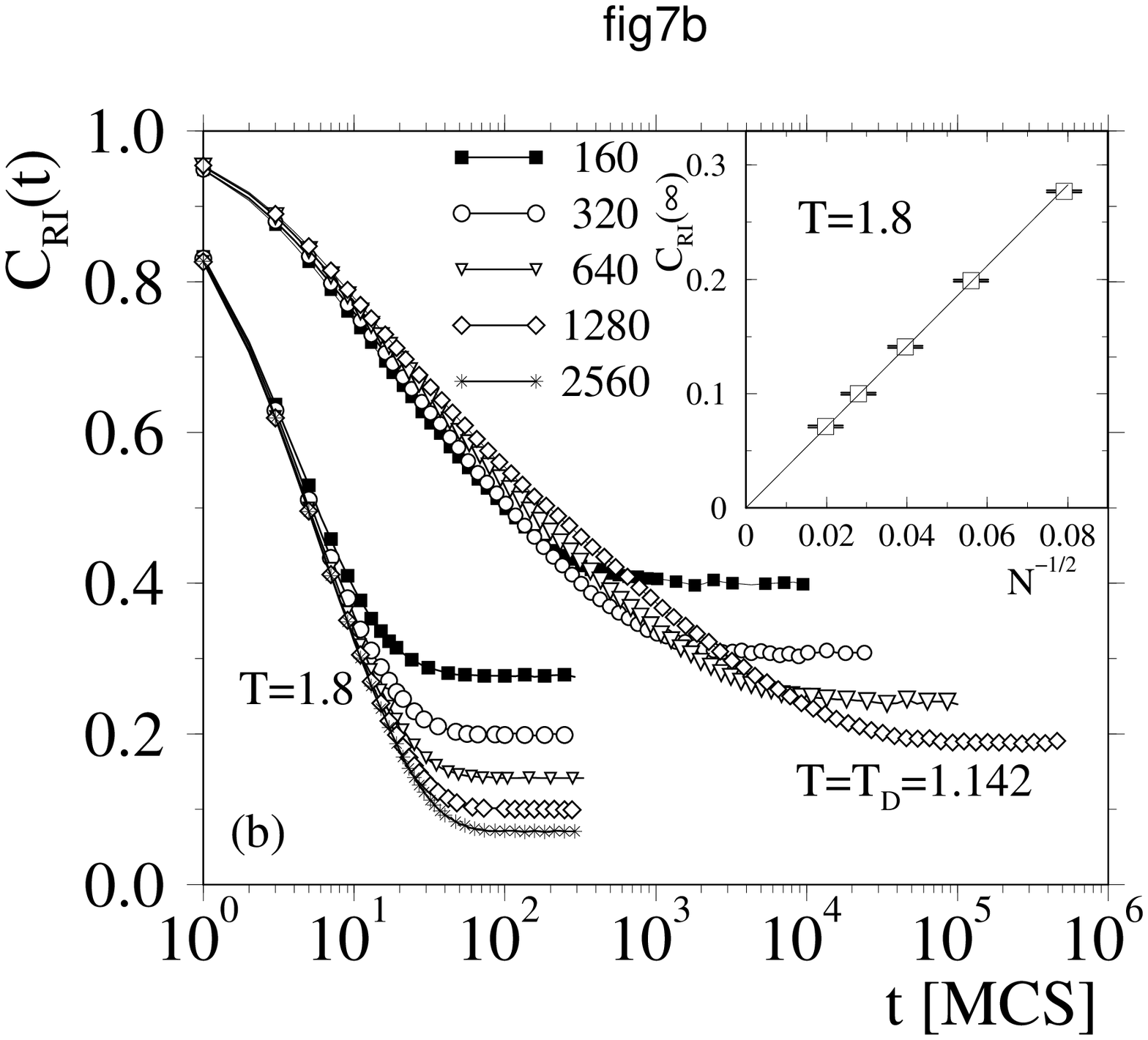,width=9.0cm,height=8.5cm}
\psfig{figure=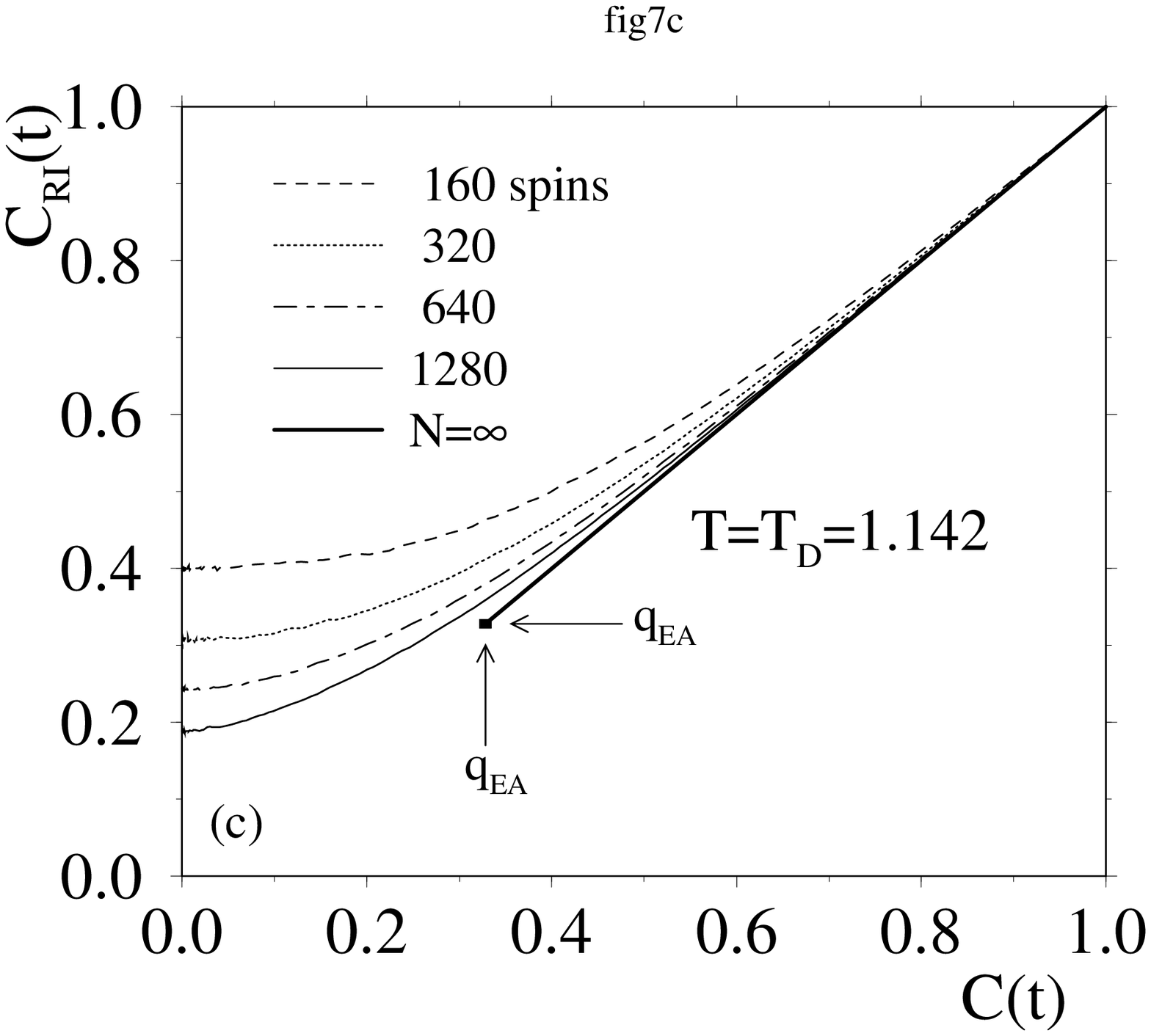,width=9.0cm,height=8.5cm}

\caption{(a) Time dependence of the correlation function $C(t)$ for $T=1.8$ and
for $ T=T_{D}=1.142$ for several values of $N$. The solid line is the
theoretical value of the Edwards-Anderson order parameter $q_{EA}(T_D)$
for $N\rightarrow \infty$ \protect\cite{DeSantis}. The dashed line shows the
value we use to define the relaxation time $\tau$. From Brangian
{\it et al.} \protect\cite{Brangian}.
(b) Time dependence of the rotationally invariant correlation function
$C_{RI}(t)$ for $T=1.8$ and for $T=T_D=1.142$ for several values of
$N$. The inset shows the limiting value $C_{RI}(t\rightarrow
\infty )$ as a function of  $N^{-1/2}$ for $T=1.8$.
(c) Parametric plot of $C_{RI}(t)$ vs. $C(t)$ at $T=T_D$ for different
values of $N$. The square indicates the plateau value obtained for
$N\rightarrow \infty $. The bold straight line describes the relation
$C_{RI}(t)=C(t)$, believed to hold for $N\rightarrow \infty $.}
\label{fig7}
\end{figure}

\begin{figure}[h]
\psfig{figure=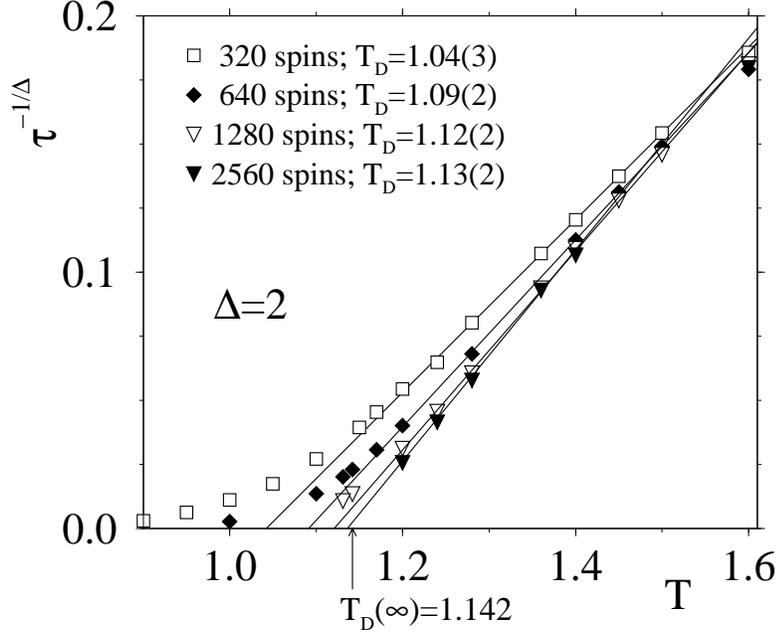,width=10.0cm,height=9.5cm}

\caption{Temperature dependence of $\tau^{-1/\Delta}$ for different system
sizes, using $\Delta =2.0$ as a trial value. The bold straight lines are
fits on a proper subset of point. The resulting extrapolated values for
$T_D(N)$ are quoted in the figure.}
\label{fig8}
\end{figure}

\begin{figure}[h]
\psfig{figure=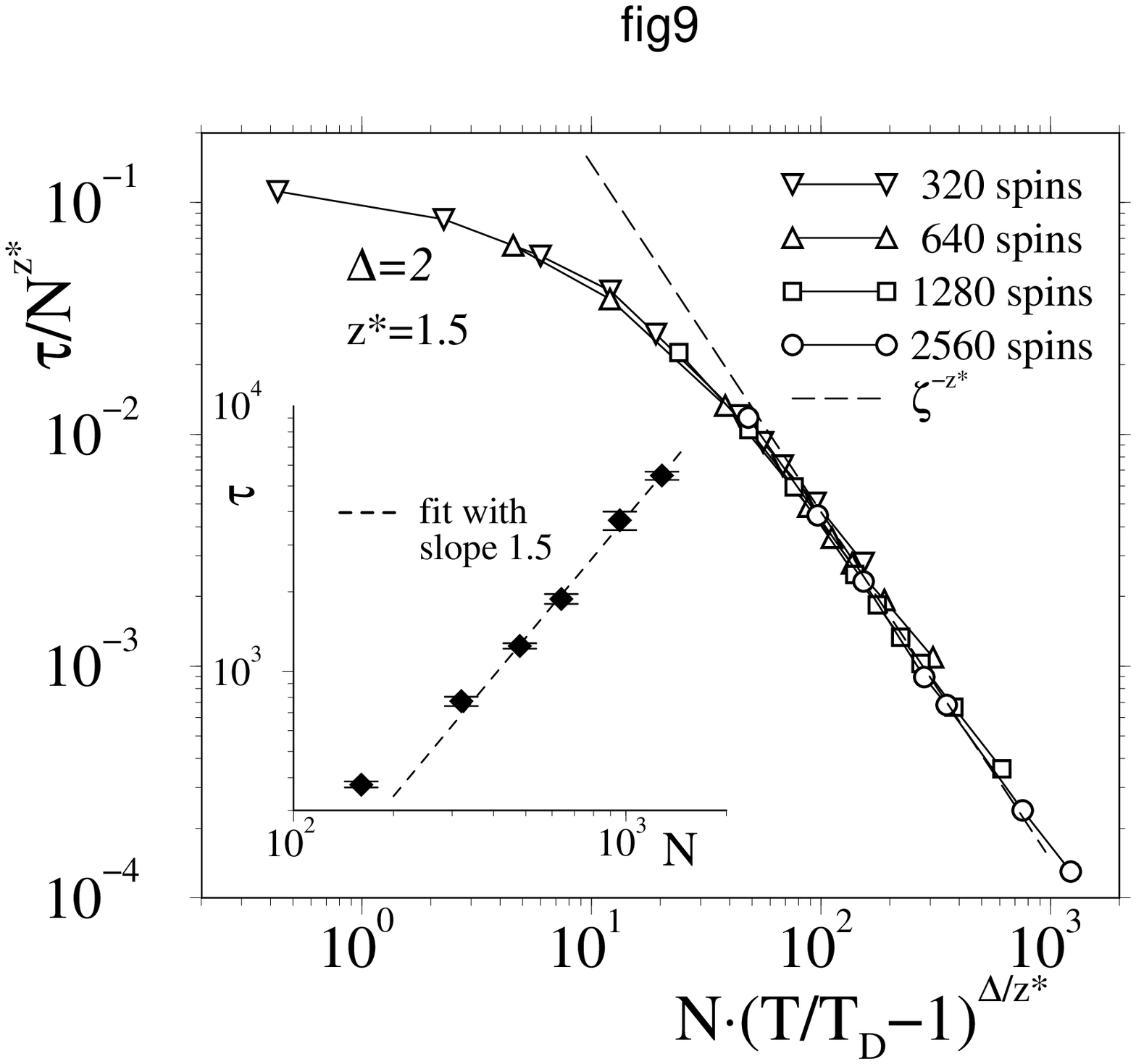,width=10.0cm,height=9.5cm}

\caption{Log-log plot of the scaled relaxation time $\tau /N^{z\ast}$
vs the scaled distance $N(T/T_{D}-1)^{\Delta /z\ast }$ from the dynamical
transition temperature $T_D$, choosing $z^{\ast }=1.5$ and
$\Delta /z^{\ast}=1.3$. The inset is a log-log plot of $\tau (T=T_D)$ vs $N$.}
\label{fig9}
\end{figure}

\begin{figure}[h]
\psfig{figure=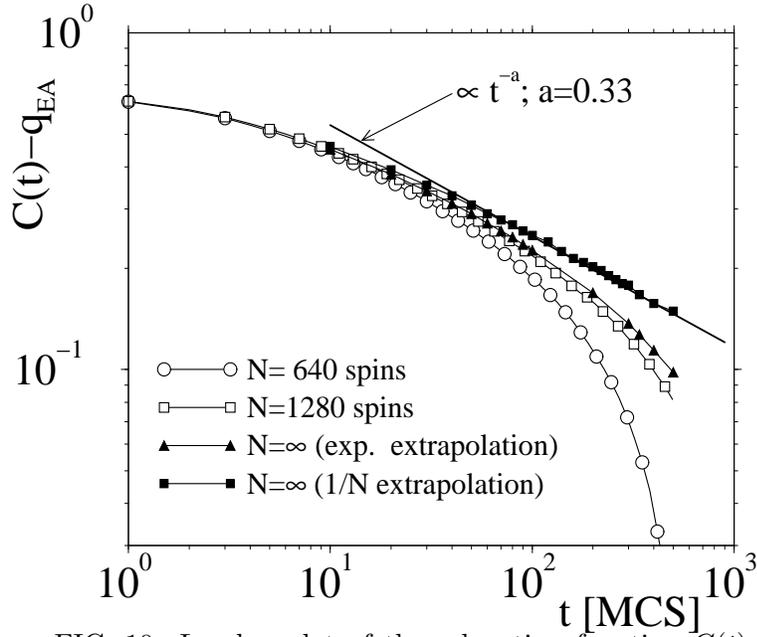,width=10.0cm,height=9.5cm}
\vspace*{-4mm}

\caption{Log-log plot of the relaxation function $C(t)-q_{EA}$ versus time
for $T=T_{D}$, using the theoretical value of $q_{EA}$, Eq.~(\ref{eq29}).
The curves with the open symbols are the data from the simulation
for two system sizes. The two curves with the filled symbols are the
extrapolation of the simulation data to the case $N=\infty$ (see main
text for details). The bold solid line is a fit with a power law.}
\label{fig10}
\end{figure}

\begin{figure}[h]
\psfig{figure=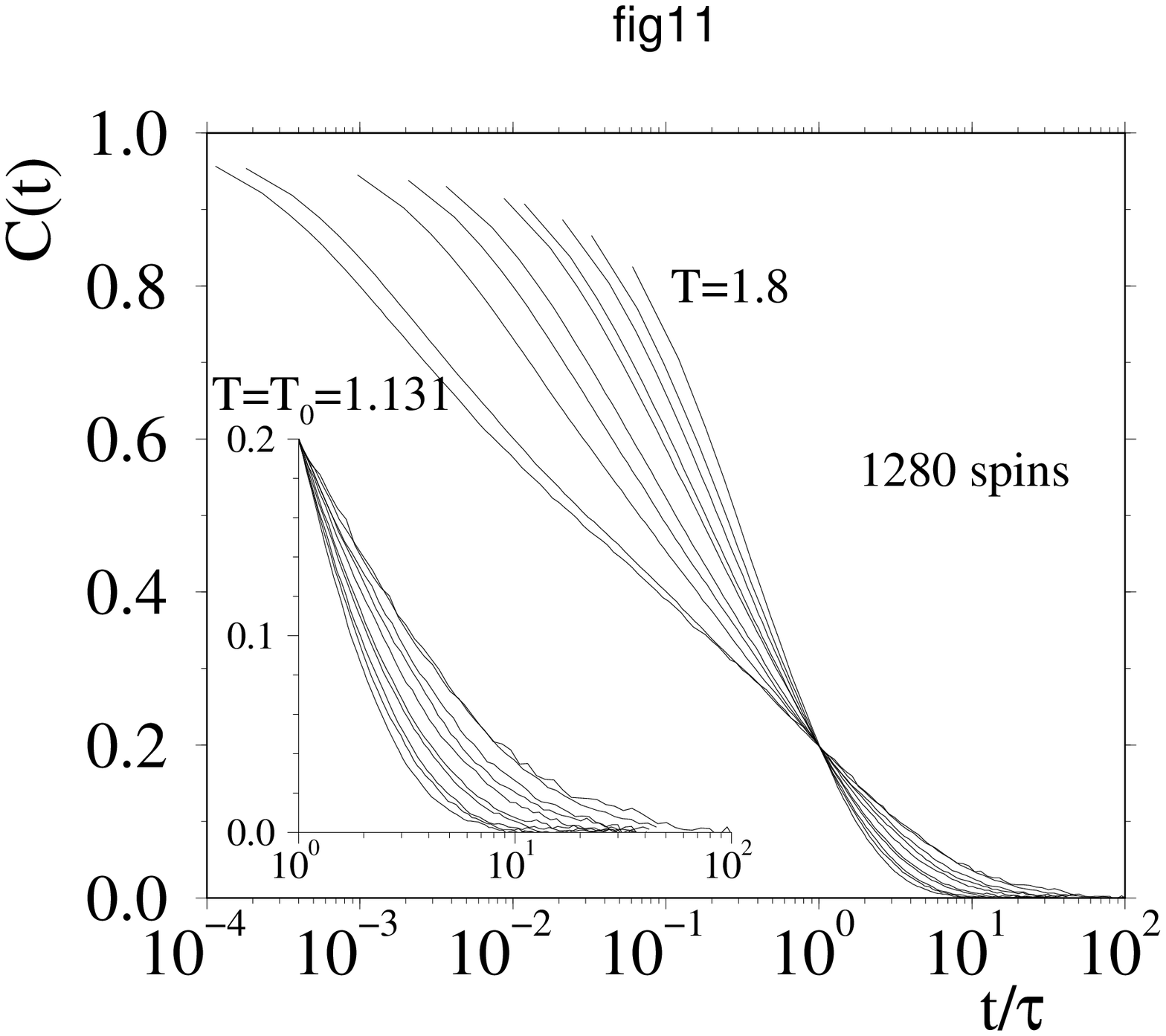,width=10.0cm,height=9.5cm}

\caption{Plot of $C(t)$ vs. $t/\tau$ (where $\tau$ is defined via
$C(t=\tau )=0.2$, cf. Eq.~(\ref{eq30})), for $N=1280$. Temperatures from
right to left:$T=1.8$, 1.6, 1.5, 1.4, 1.360, 1.280, 1.240, 1.2, 1.142,
and 1.131.  The inset shows a magnification of the part of the curves
for $t/\tau >1$.}
\label{fig11}
\end{figure}

\begin{figure}[h]
\psfig{figure=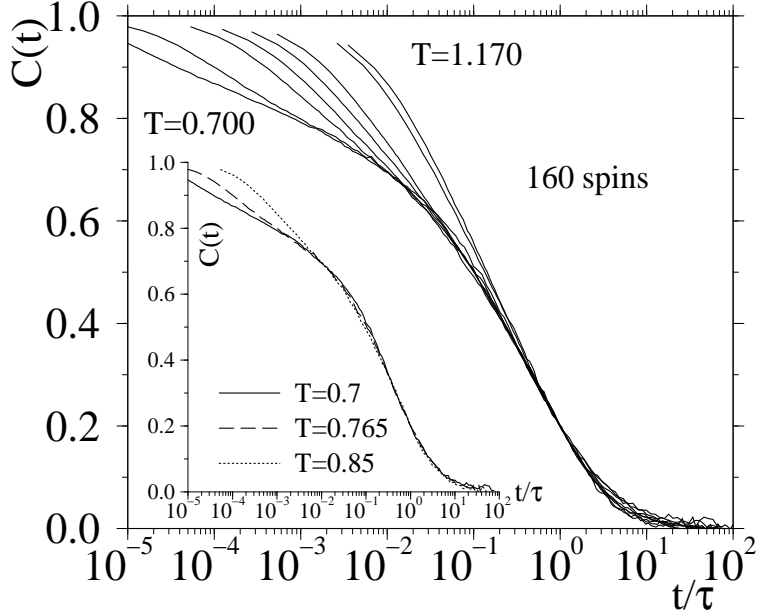,width=10.0cm,height=9.5cm}

\caption{Plot of $C(t)$ vs. $t/\protect\tau $ for $N=160$. The temperatures
are $T=0.7$, 0.765, 0.85, 0.9, 0.95, 1.0, 1.142, and 1.17 (left to right).
The inset shows the same data, but including only the three lowest temperatures.}
\label{fig12}
\end{figure}

Gratifyingly, we have established that the Monte Carlo results are
qualitatively compatible with the theoretical predictions, although the
finite size effects found were unexpectedly strong (i.e. they occur over a
very wide temperature range as well) and they are not understood in detail,
and hence we have found it too difficult to extract the various temperatures
mentioned above directly from the simulation itself. E. g., for the sizes
available, the method of looking for intersection points of the fourth \
oder cumulant or the Guerra parameter do not allow for a reliable estimation
of $T_0$ (Fig.~\ref{fig4}). Similarly, one could estimate the temperatures 
$T_s$ (Fig.~\ref{fig2}), $T_K$ (Fig.~\ref{fig1}b) and $T_D$\ (Fig.~\ref
{fig8}) from a naive analysis of the data only very roughly. However,
if one uses the theoretical knowledge on $T_D$, one can estimate both
the exponent $\Delta $ mentioned above and the exponent $z^{\ast }$ for
the size dependence of the time $\tau $ at $T_D$\ $\left( \tau \propto
N^{z^{\ast }}\right) $ from a dynamical finite size scaling analysis
(Fig.~\ref{fig9}).  We also found evidence that the predicted power
law decay of the spin autocorrelation function $C(t)-q_{EA}
\propto t^{-a}$ occurs for $T\approx T_D$ (Fig.~\ref{fig10}), but we
could not confirm the expected time-temperature superposition principle
(Fig.~\ref{fig11}). We also analysed the modified disconnected cumulants
proposed in reference \cite{Picco:2001}, which should give a better
estimation of the spin glass transition temperature in systems exibiting
one step replica symmetry breaking patterns, but with the system sizes at
our disposal they do not seem to work better than the corresponding
connected parameters \cite{brangian_phd}.

Also some steps were taken to analyze the dynamics for $T\leq T_D$,
by investigating the relaxation function $C_{i}\left( t\right) $ of
individual spins and corresponding relaxation times 
(Figs.~\ref {fig14}-\ref{fig17}). We find that the reason for the observed
non-exponential relaxation of the {\it mean} relaxation function
$C(t)$ is related to the presence of a very strong dynamical
heterogeneity. Furthermore we found that certain spins form dynamical
clusters, the reason for which are likely their strong bonds between
them. However, this mechanism seems not to be the only one and hence
this point has to be investigated in the future in more detail.

Thus, although many exact results are known for this model, and - unlike
in models of the structural glass transition - we can equilibrate the
system also at temperatures significanly below $T_D$ for a range of
sizes $\left( N\leq 640\right)$, we still are not able to answer many
questions. Nevertheless, we think that the present model is a prototype
model for glass transitions, and if better simulation algorithms become
available, further studies of the present model should be very rewarding.

{\bf Acknowledgements:} One of us (C.B) was partially supported by
the Deutsche Forschungsgemeinschaft, Sonderforschungsbereich 262/D1. We
thank F. Ritort for information on the energy vs. temperature curve 
(Fig.~\ref{fig1}a). We are grateful to the John von Neumann Institute for
Computing (NIC J\"{u}lich) for a generous grant of computer time at the CRAY
T3E.

\vspace*{-12mm}

\vspace*{-3mm}

\vspace*{-9mm}

\begin{figure}[]
\psfig{figure=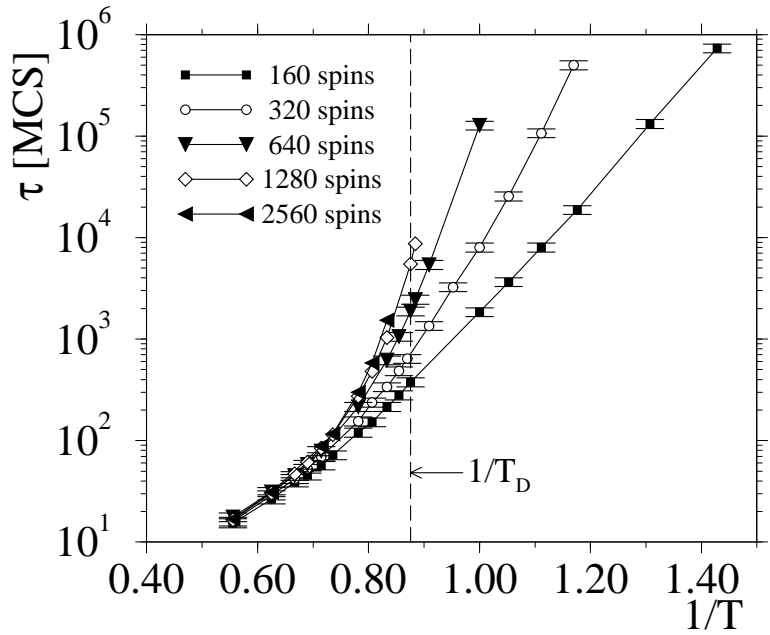,width=10.0cm,height=9.5cm}

\caption{Relaxation time $\tau$ plotted vs. inverse temperature for
different system sizes. The broken vertical line indicates the location
of the dynamical transition. Note the choice of a logarithmic scale for
the ordinate. Error bars of $\tau$ are mostly due to the sample-to-sample
fluctuation.}
\label{fig13}
\end{figure}

\begin{figure}[h]
\psfig{figure=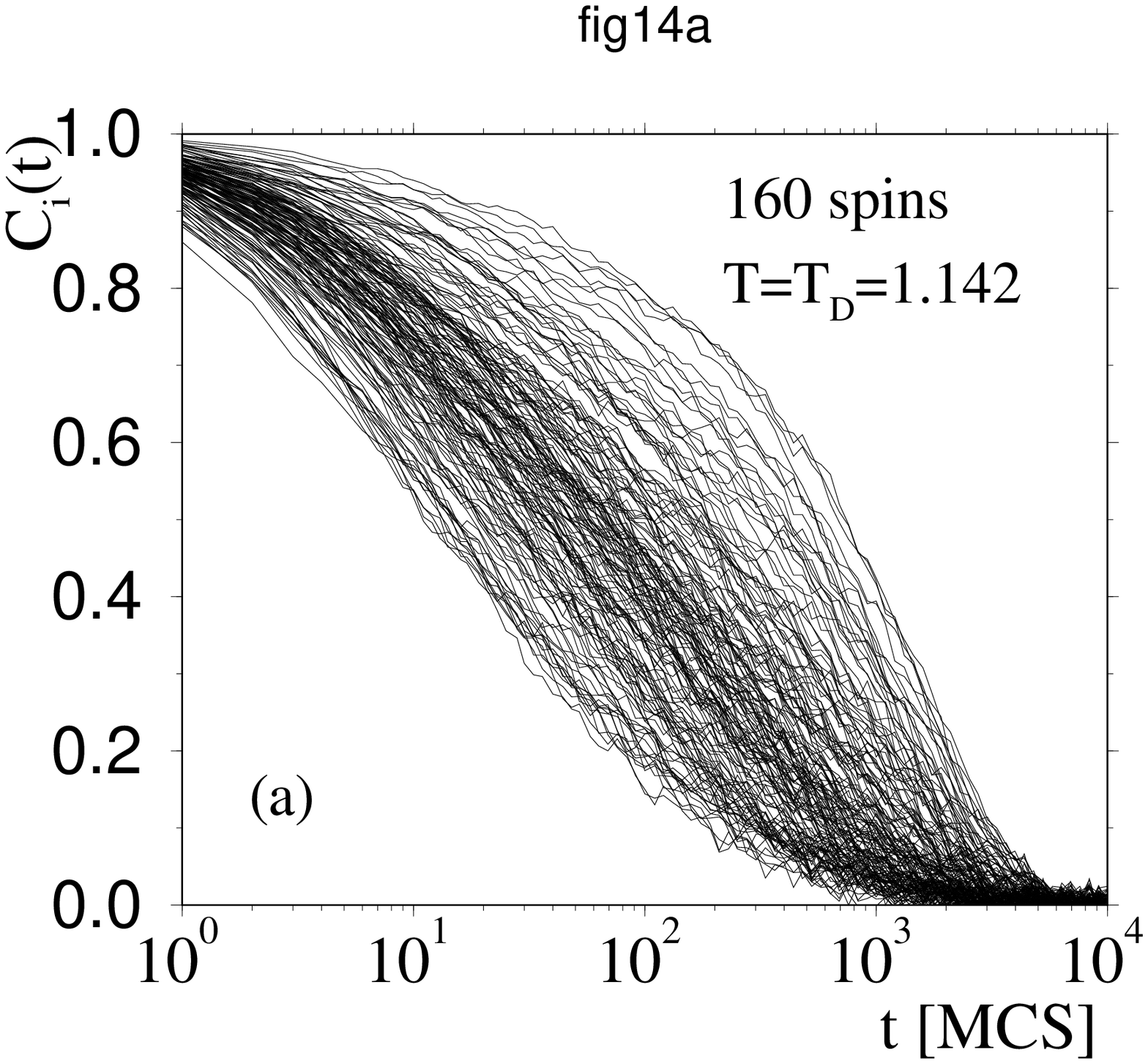,width=10.0cm,height=9.5cm}
\psfig{figure=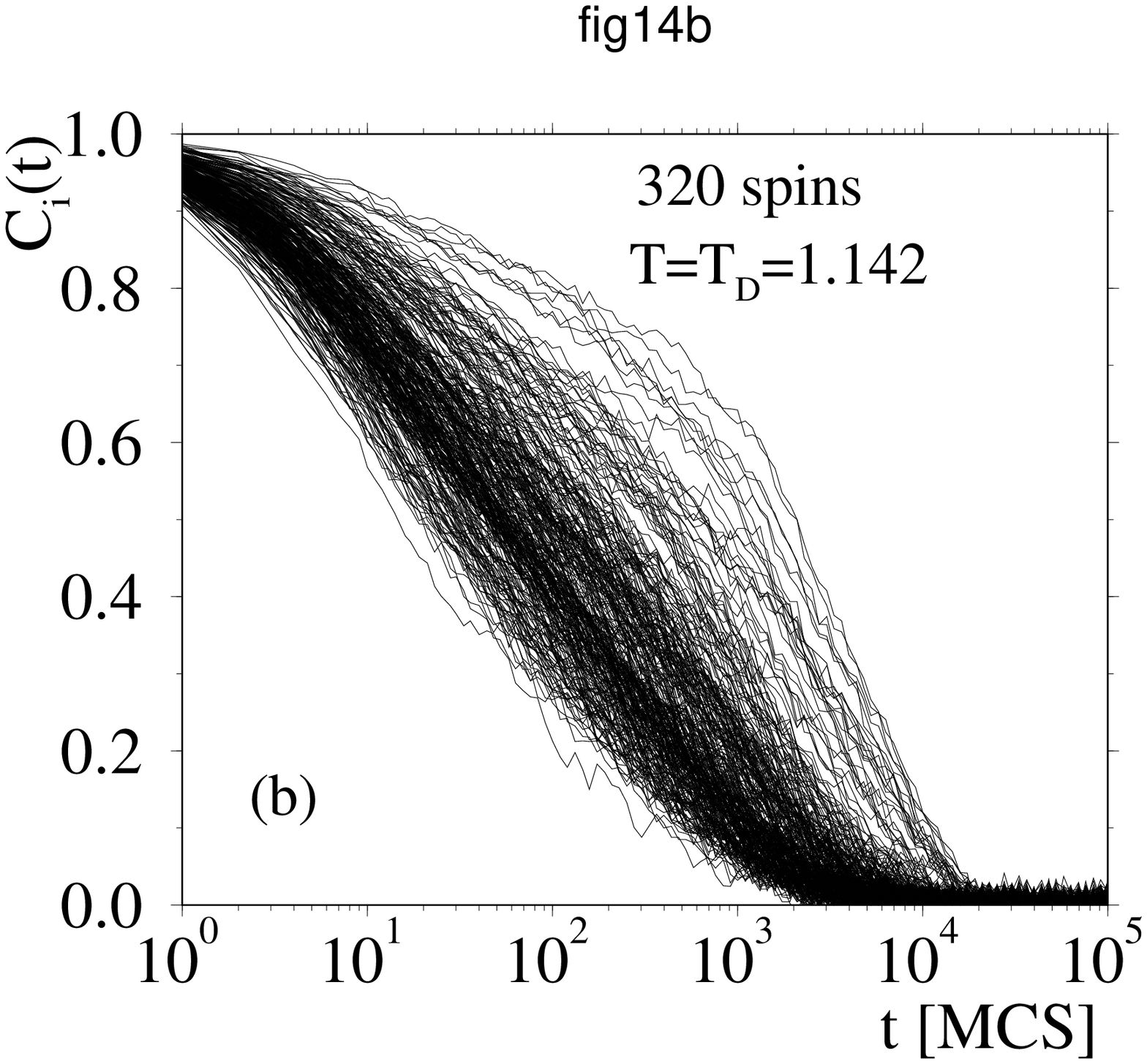,width=10.0cm,height=9.5cm}
\psfig{figure=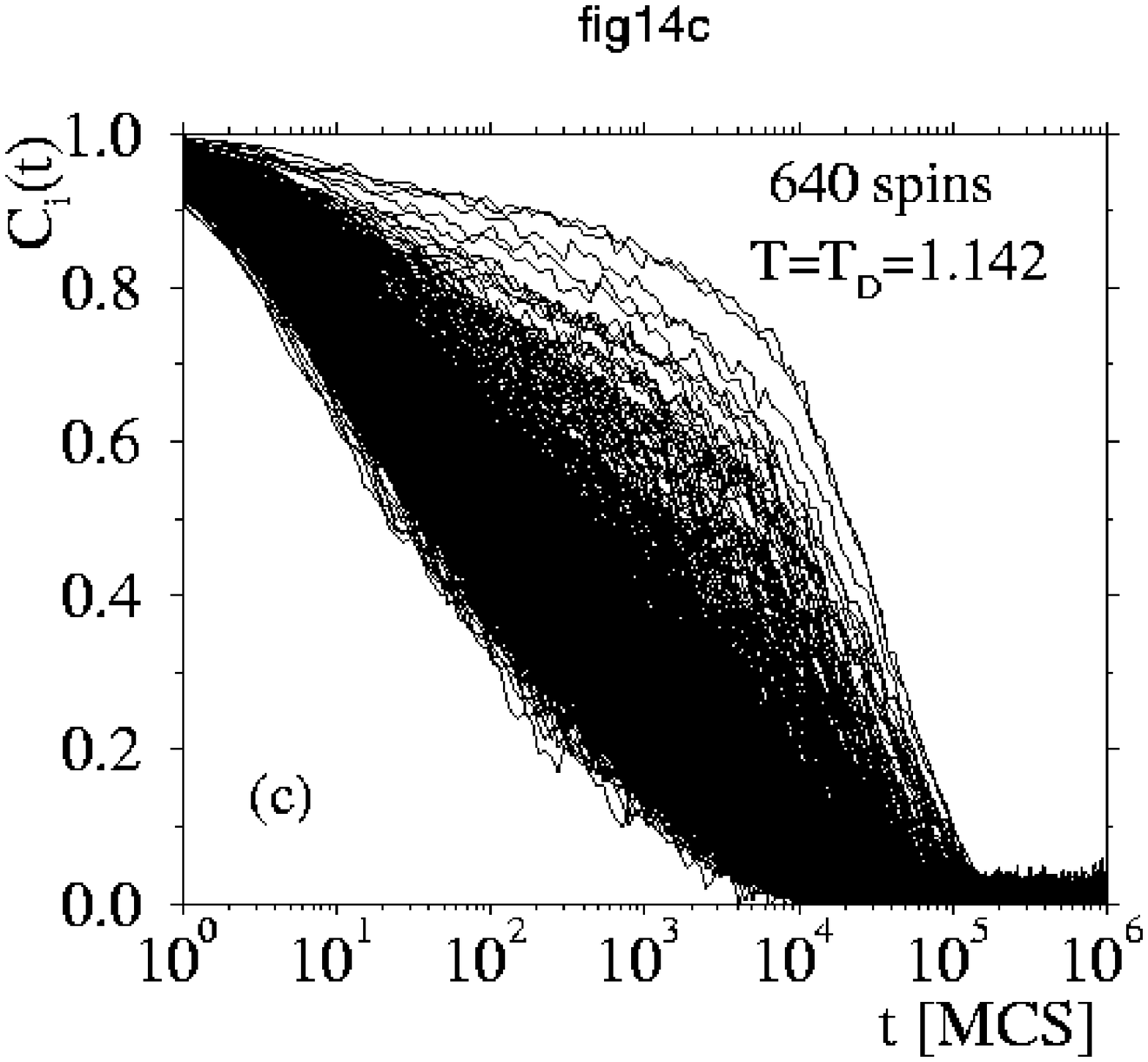,width=12.0cm,height=9.5cm}

\caption{Time dependence of the single spin autocorrelation function
$C_{i}(t)$ at $T=T_{D}$ for $N=160 $(a), $N=320$ (b), and $N=640$
(c). Each of the curves corresponds to a different spin.}
\label{fig14}
\end{figure}

\begin{figure}[h]
\psfig{figure=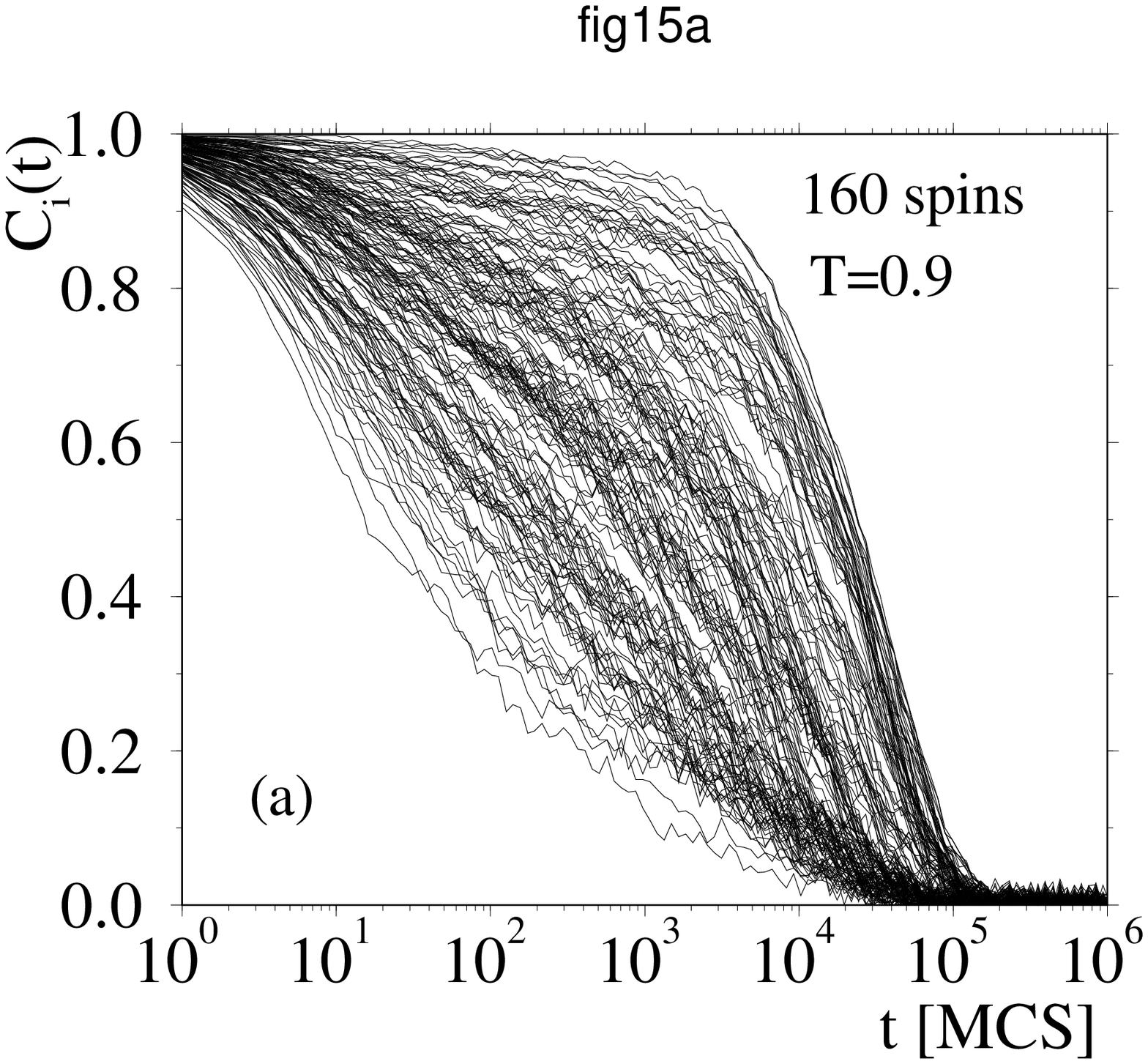,width=10.0cm,height=9.5cm}
\psfig{figure=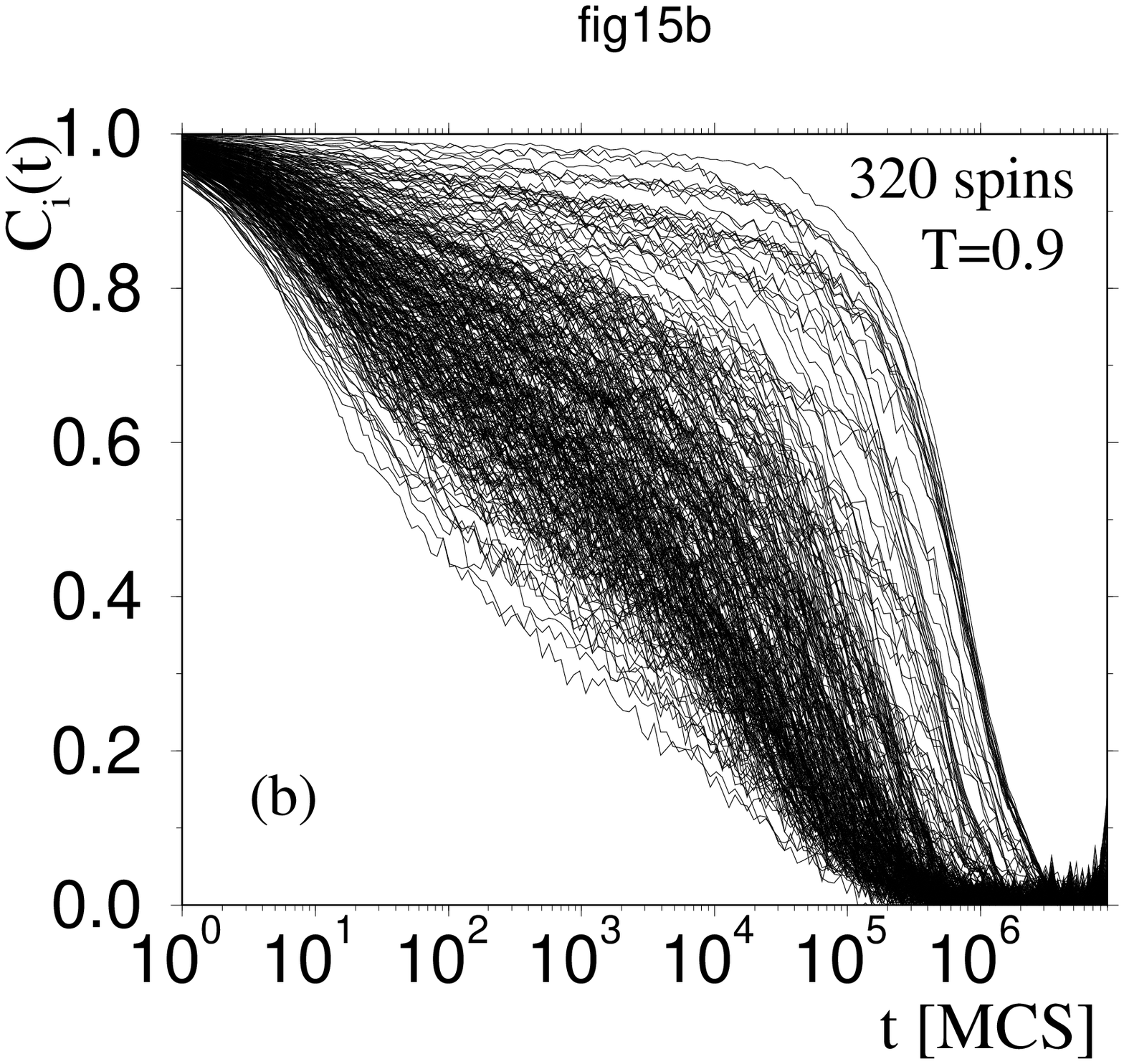,width=10.0cm,height=9.5cm}

\caption{Time dependence of the single spin autocorrelation function
$C_{i}(t)$ at $T=0.9$ for $N=160$ (a) and for $N=320$ (b). Each of the curves
corresponds to a different spin.}
\label{fig15}
\end{figure}

\begin{figure}[h]
\psfig{figure=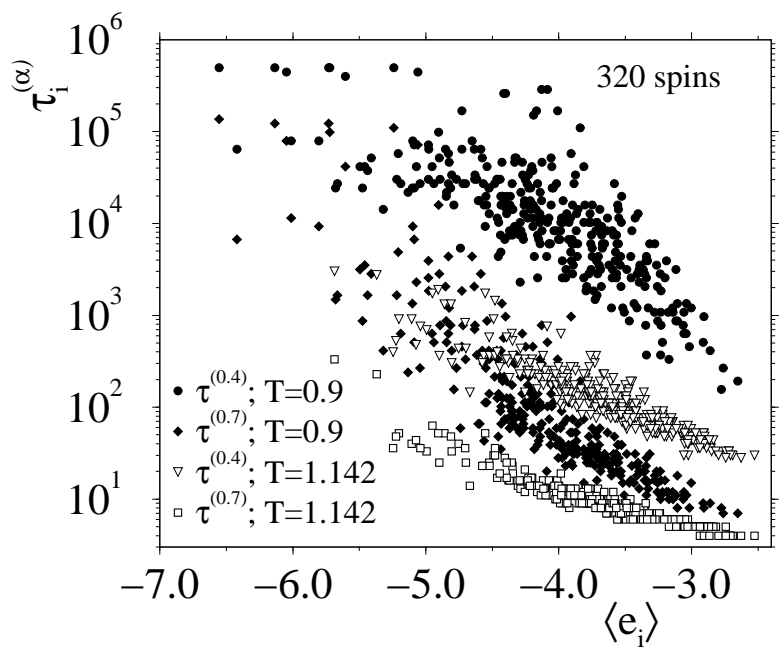,width=10.0cm,height=9.5cm}

\caption{Scatter plot of $\tau_{i}$, the mean relaxation time of spin
$i$ as defined in Eq.~(\ref{eq36}), versus the mean energy $\langle
e_{i}\rangle$. The open and closed symbols correspond to $T=1.142$ and
$T=0.9$, respectively. The points are for a typical sample of
320 spins.}
\label{fig16}
\end{figure}

\begin{figure}[h]
\psfig{figure=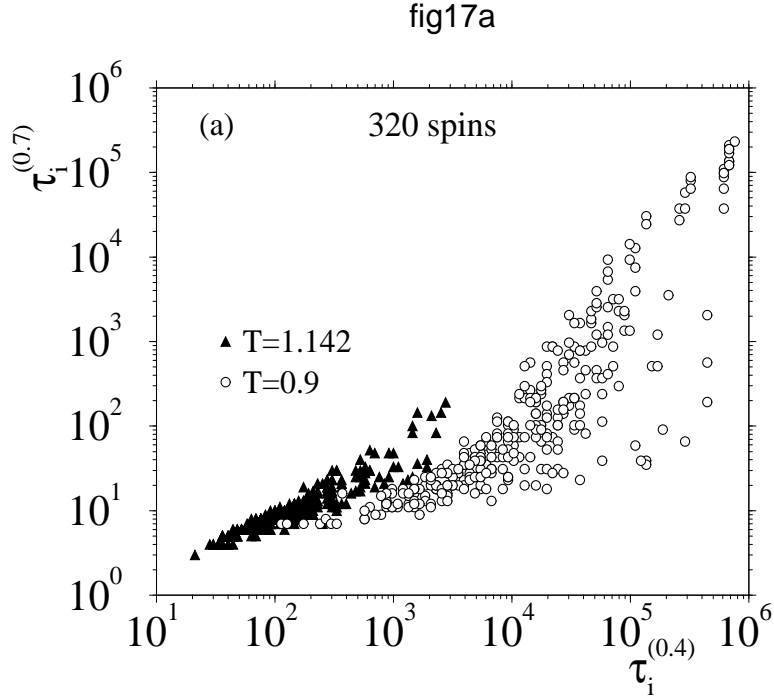,width=10.0cm,height=9.5cm}
\psfig{figure=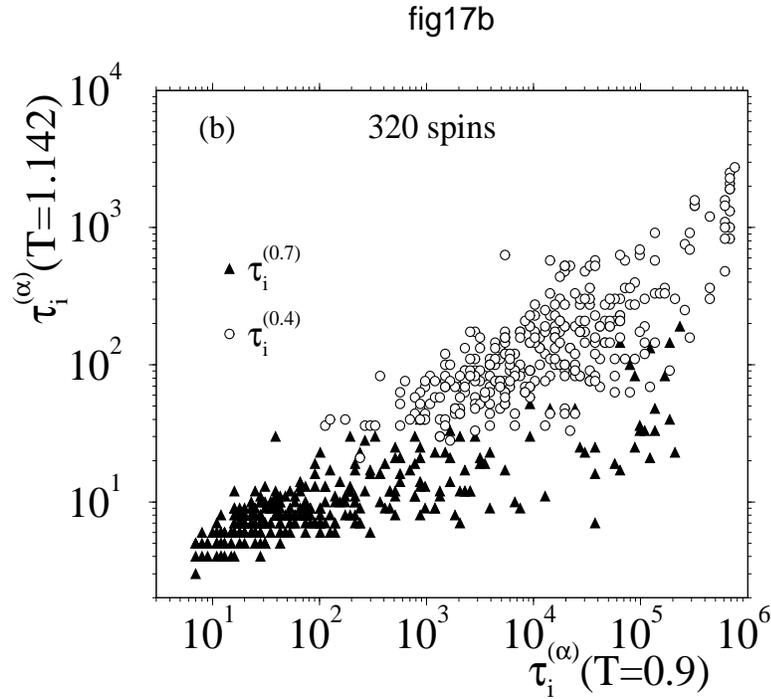,width=10.0cm,height=9.5cm}

\caption{a) Plot of $\tau_i^{(0.7)}$ [defined by $C_{i}(t=\tau_i)=0.7$] versus
$\tau_i^{(0.4)}$ [defined by $C_{i}(t=\tau_i)=0.4$], for $T=0.9$ and $T=1.142$
(open and filled symbols, respectively) showing that the two relxation times are
correlated. Each point correspond to a different spin.
b) Plot of the relaxation times $\tau_i^{(0.4)}$ and $\tau_i^{(0.7)}$ at $T=1.142$
versus these relaxation times at $T=0.9$. Each point correspond to a different
spin.}
\label{fig17}
\end{figure}

\begin{figure}[h]
\psfig{figure=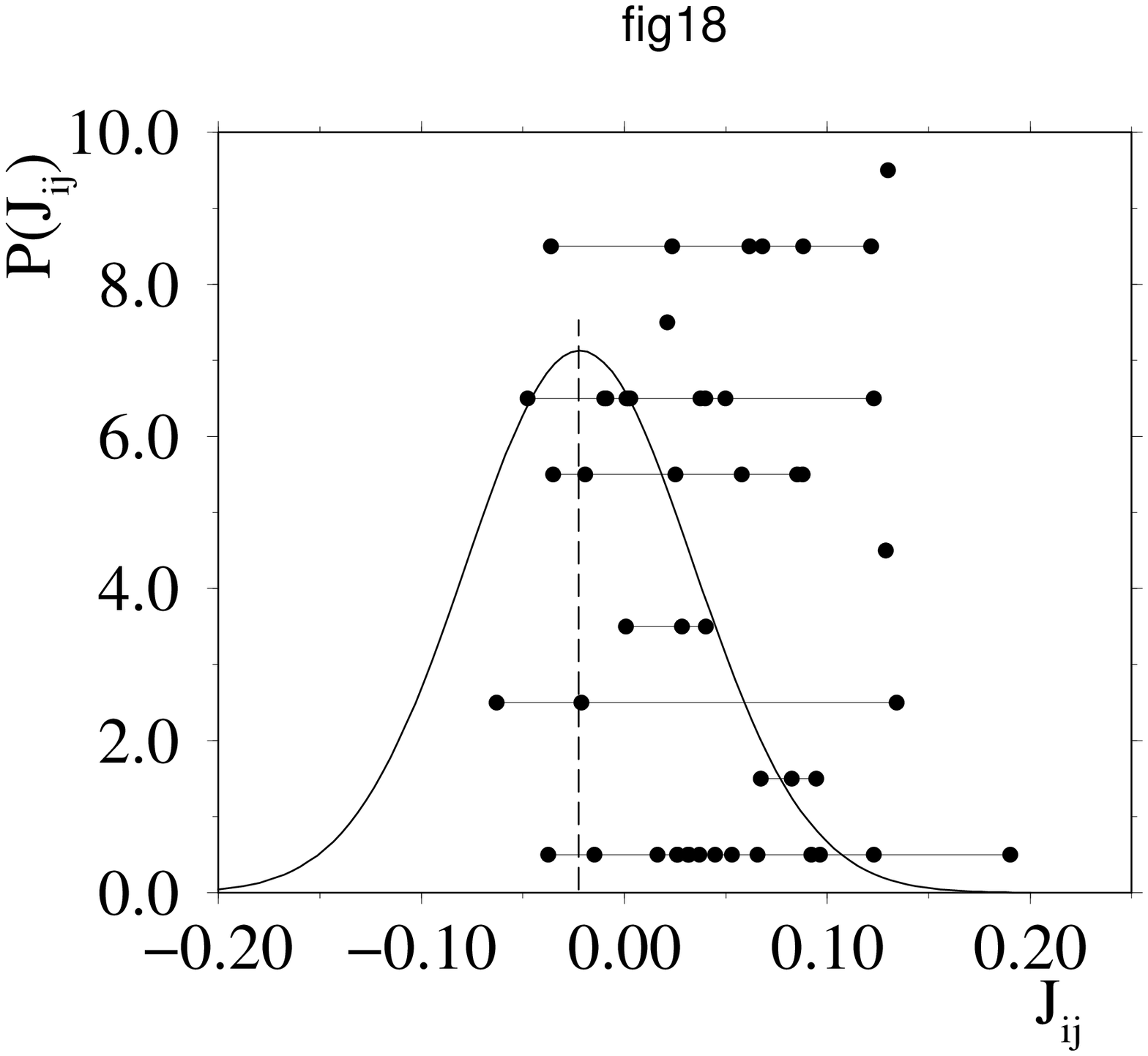,width=10.0cm,height=9.5cm}

\caption{Values of the bonds between spins with very slow relaxation in 10
different disorder realizations for $320$ spins at a temperature $T=0.9$
(filled circles). For clarity the points have been displaced vertically
by various amounts. The continuous curve shows the Gaussian distribution
from which the $J_{ij}$ are extracted and the vertical dashed line shows
its mean.}
\label{fig18}
\end{figure}


\begin{references}

\bibitem{Monasson:1995} 
Monasson R 1995
{\it Phys. Rev. Lett.} \textbf{75} 2847

\bibitem{Bouchaud:1998}
Bouchaud J-P, Cugliandolo L, Kurchan J, and M\'ezard M 1998
in \textit{Spin Glasses and Random Fields}, ed. A P Young, World
Scientific (Singapore) 

\bibitem{Franz:1998}
Franz S and Parisi G 1998
{\it Physica A} \textbf{261} 317

\bibitem{Mezard:1999a}
M\'ezard M and Parisi G 1999
{\it Phys. Rev. Lett.} \textbf{82} 747

\bibitem{Kirkpatrick}  
Kirkpatrick T R and Wolynes P G 1987 
{\it Phys. Rev.} B {\bf 36} 8552;
%
Kirkpatrick T R and Thirumalai D 1988 
{\it Phys. Rev.} B {\bf 37 } 5342;
%
Thirumalai D and Kirkpatrick T R 1988 
{\it Phys. Rev}. B {\bf 38 } 4881

\bibitem{Gotze}  
G\"{o}tze W 1990 
in {\it Liquids, Freezing and the Glass
Transition}, Hansen J P, Levesque D and Zinn-Justin J, eds. (Amsterdam,
North Holland) p. 287;
%
{\it J. Phys.: Condens. Matter} {\bf 11} A1

\bibitem{Elderfield}  
Elderfield D and Sherrington D 1983 
{\it J. Phys. C: Solid State Phys}. {\bf 16} L491, L971, L1169

\bibitem{Erzan} 
Erzan A and Lage E J S 1983 
{\it J. Phys. C: Solid State Phys}. {\bf 16} L55

\bibitem{Gross}  
Gross D J, Kanter I and Sompolinsky H 1985 
{\it Phys. Rev. Lett.} {\bf 55} 304

\bibitem{Carmesin} 
Carmesin H O and Binder K 1988 
{\it J.Phys.A:Math.Gen.} {\bf 21} 4053

\bibitem{Cwilich} 
Cwilich G and Kirkpatrick T R 1989 
{\it J. Phys. A: Math. Gen.} {\bf 22} 4971;
%
Cwilich G 1990 
{\it J. Phys. A: Math. Gen.} A {\bf 23} 5029

\bibitem{Scheucher}  
Scheucher M, Reger J D, Binder K and Young A P 1990 
{\it Phys.  Rev.} B {\bf 42} 6881;
%
{\it Europhys. Lett.} {\bf 14}, 119;
%
Scheucher M and Reger J D 1993 
{\it Z. Phys.} B {\bf 91} 383

\bibitem{DeSantis}  
De Santis E, Parisi G and Ritort F 1995 
{\it J. Phys. A: Math.  Gen}. {\bf 28} 3025

\bibitem{Schreider}  
Schreider G and Reger J D 1995 
{\it J. Phys. A: Math. Gen.} {\bf 28} 317

\bibitem{Rehul}  
Reuhl M, Nielaba P and Binder K 1999 
{\it Eur. Phys. J.} B {\bf 2} 225

\bibitem{Dillmann}  
Dillmann O, Janke W and Binder K 1998 
{\it J. Stat. Phys.} {\bf 92} 57

\bibitem{Lobe}  
Lobe B, Janke W and Binder K 1999 
{\it Eur. Phys. J.} B {\bf 7} 289

\bibitem{Marinari}
Marinari E, Mossa S and Parisi G 1999
{\it Phys. Rev. }B \textbf{59}, 8401

\bibitem{Hukushima:2000} 
Hukushima K and Kawamura H 2000 
{\it Phys. Rev.} E {\bf 62}, 3362


\bibitem{Edwards}  
Edwards S F and Anderson P W 1975 
{\it J. Phys. F: Metal Physics} {\bf 5} 965

\bibitem{Sherrington}  
Sherrington D and Kirkpatrick S 1975 
{\it Phys. Rev. Lett.} {\bf 35}, 1972

\bibitem{Binder:1986}  
Binder K and Young A P 1986 
{\it Rev. Mod. Phys.} {\bf 58} 801

\bibitem{Mezard:1987}  
M\'ezard M, Parisi G and Virasoro M A 1987 
{\it Spin Glass Theory and Beyond} (Singapore, World Scientific)

\bibitem{Fischer}  
Fischer K H and Hertz J A 1991 
{\it Spin Glasses} (Singapore, World Scientific)

\bibitem{Stein}  
Stein D S 1992 
{\it Spin Glasses and Biology} (Singapore, World Scientific)

\bibitem{Parisi:1992}  
Parisi G 1992 
{\it Field Theory, Disorder and Simulations} (Singapore, World Scientific)

\bibitem{Young}
Young A P (ed.) 1998 
{\it Spin Glasses and Random Fields} (Singapore, World Scientific)

\bibitem{Potts}  
Potts R B 1952 
{\it Proc. Camb. Phil. Soc.} {\bf 48} 106

\bibitem{Wu}  
Wu F Y 1982 
{\it Rev. Mod. Phys.} {\bf 54} 235; {\it ibid.} {\bf 55} 315

\bibitem{Zia}  
Zia R K P and Wallace D J 1975 
{\it J. Phys. A.: Math. Gen.} {\bf 8} 1495

\bibitem{Kauzmann}  
Kauzmann W 1948 
{\it Chem. Rev}. {\bf 43} 219

\bibitem{Gibbs}  
Gibbs J H and DiMarzio E A 1958 
{\it J. Chem. Phys.} {\bf 28}
373

\bibitem{Jackle}  
J\"{a}ckle J 1986 
{\it Rep. Progr. Phys.} {\bf 49} 171

\bibitem{Parisi:1987}  
Parisi G 1987 
in {\it Complex Behavior of Glassy Systems} Rubi M 
and Perez-Vicente, eds. (Berlin, Springer) p. 111

\bibitem{Binder:1999}  
Binder K, Baschnagel J, Kob W and Paul W 1999 
{\it Physics\ World } {\bf 12}, 54

\bibitem{coluzzi99}
Coluzzi B, Verrocchio P, M\'ezard M and Parisi G 1999
{\it J. Chem. Phys.} \textbf{111} 9039 

\bibitem{sciortino99}
Sciortino F, Kob W and Tartaglia P 1999
{\it Phys. Rev. Lett.} {\bf 83} 3214

\bibitem{Crisanti}  
Crisanti A and Ritort F 2000 
{\it Europhys. Lett.} {\bf 51} 147

\bibitem{Kob:1999}  
Kob W 1999 
{\it J. Phys.: Condens. Matter} {\bf 11} R85

\bibitem{Mezard:1999b}  
M\'ezard M 1999 
{\it Physica} A {\bf 265} 359

\bibitem{Parisi:2000}  
Parisi G 2000 
{\it Physica} A {\bf 280} 115


\bibitem{Hochli}  
H\"{o}chli U T, Knorr K and Loidl A 1990 
{\it Adv. Phys}. {\bf 39} 405

\bibitem{Binder:1992a}  
Binder K and Reger J D 1992 
{\it Adv. Phys.} {\bf 41} 547

\bibitem{Binder:1998}  
Binder K in Ref. \cite{Young} p. 99

\bibitem{Sillescu}  
Sillescu H 1999 
{\it J. Noncryst. Solids} {\bf 243} 81

\bibitem{ediger00}
Ediger M D 2000
{\it Annu. Rev. Phys. Chem.} {\bf 51} 99 

\bibitem{Guerra}  
Guerra F 1996 
{\it Int. J. Mod. Phys}. B{\bf 10} 1675
%
Marinari E et \textit{al}. 1998
{\it Phys. Rev. Lett.} {\bf 81} 1698

\bibitem{Binder:1997}  
Binder 1997 
{\it Rep. Progr. Phys.} {\bf 60} 487; 
%
Landau D P and Binder K 2000 
{\it A Guide to Monte Carlo simulations in statistical Physics}
(Cambridge, Cambridge Univ. Press)

\bibitem{Hukushima:1996}  
Hukushima K and Nemoto K 1996 
{\it J. Phys. Soc. Jpn}. {\bf 64} 1604

\bibitem{Kob:2000}  
Kob W, Brangian C, St\"{u}hn T and Yamamoto R 2000 
in {\it Computer Simulation Studies in Condensed Matter Physics XIII} Landau D P,
Lewis S P and Sch\"{u}ttler H B, eds, (Berlin, Springer) p. 134

\bibitem{brangian_phd}
Brangian C 2001 
Dissertation (Johannes-Gutenberg-Universit\"{a}t Mainz, in preparation)

\bibitem{Binder:1981}
Binder 1981
{\it Z. Phys. B} {\bf 45} 61

\bibitem{Ritort_comm}  
Ritort F, 
{\it private communication}

\bibitem{Wolfgardt}  
Wolfgardt M, Baschnagel J, Paul W and Binder K 1996 
{\it Phys.  Rev}. E {\bf 54} 1535

\bibitem{Binder:1992b}  
Binder\ K 1992 
in {\it Computational Methods in Field Theory},
Gausterer H and Lang C B, eds,  (Berlin, Springer) p. 59

\bibitem{Vollmayr}  
Vollmayr K, Reger J D, Scheucher M and Binder K 1993 
{\it Z.  Phys.} B {\bf 91} 113

\bibitem{Brangian}  
Brangian C, Kob W, and Binder K 2001 
{\it Europhys. Lett.} {\bf 53} 756

\bibitem{Suzuki}  
Suzuki M 1977 
{\it Progr. Theor. Phys.} {\bf 58} 1142

\bibitem{Hohenberg}  
Hohenberg P C and Halperin B I 1977 
{\it Rev. Mod. Phys.} {\bf 49} 435

\bibitem{Bhatt}  
Bhatt R N and Young A P 1992 
{\it Europhys. Lett.} {\bf 20} 59

\bibitem{Wortis}
Jayaprakash C, Chalupa J and  Wortis M 1977
{\it Phys. Rev. B} \textbf{15} 1495

\bibitem{mackenzie}
Mackenzie N D and Young A P 1983
{\it J. Phys.} C \textbf{16} 5321 

\bibitem{Billoire}
Billoire A and Marinari E
preprint cond-mat/0101177  

\bibitem{horbach99}
Horbach J and Kob W 1999 
{\it Phys. Rev.} B {\bf 60} 3169

\bibitem{glotzer98}
Glotzer S C, Jan N and Poole P H 1998
{\it Phys. Rev. E} {\bf 57} 7350

\bibitem{kondor89}
Kondor I 1989
{\it J. Phys. A: Math. Gen.} {\bf 22} L163 

\bibitem{ritort94}
Ritort F 1994
{\it Phys. Rev. }B \textbf{50} 6844

\bibitem{kim}
Kim K and Yamamoto R 2000
{\it Phys. Rev.} E \textbf{61} R41 

\bibitem{biroli99}
Biroli G and Monasson R 2000
{\it Europhys. Lett.} \textbf{50} 155 

\bibitem{Parisi:1993}
Parisi G, Ritort F and Slanina F 1998
{\it J. Phys. }A \textbf{26} 247 ; 
%
1998 \textit{ibid.} \textbf{26} 3775 

\bibitem{Picco:2001}
Picco M, Ritort F and Sales M 2001
{\it Eur. Phys. J. }B {\bf 19} 565 

\end{references}
\end{document}